\documentclass{sdsu-thesis}
\usepackage{epsfig}
\usepackage{graphicx}

\usepackage{amsmath}
\usepackage{amsfonts}
\usepackage{amssymb}
\usepackage{amsthm}
\usepackage{lmodern}
\usepackage{anyfontsize}
\usepackage{tabto}
\usepackage{placeins}
\usepackage{esvect}

\usepackage[bf,labelsep=period,textfont=bf]{caption}

\usepackage{url}
\usepackage{longtable}
\usepackage[normalem]{ulem}

\usepackage{blindtext}
\usepackage{enumitem}
\usepackage{xcolor}
\usepackage{eucal}

\usepackage{datatool}
\usepackage[nonumberlist,section=paragraph]{glossaries}
\makeglossaries
\newglossaryentry{Final Model}{
  name=Final Model,
  description={The last valid configuration reported of the accreting White Dwarf before the stopping condition is met, based either on the total nuclear power generated or the maximum temperature in any zone being greater than the maximum we specify}
}

\newglossaryentry{G05}{
  name=G05,
  description={Gasques et al. \cite{G05}}
}

\newglossaryentry{mesa}{
   name=MESA,
   description={Modules for Experiments in Stellar Astrophysics.  A scientific software application for performing simulations of stellar evolution, including the evolution of accreting white dwarfs}
}

\newglossaryentry{Pycnonuclear Reactions}{
  name=Pycnonuclear Reactions,
  description={Nuclear reactions that occur at high densities, such as the densities found at the center of white dwarfs and the inner crusts of neutron stars}
}

\newglossaryentry{RMF}{
  name=RMF,
  description={Formulae used to calculate nucleon-nucleon interaction potentials derived from Relativistic Mean Field theory in the paper by Singh et al. \cite{SINGH2012}}
}

\newglossaryentry{SK}{
  name=SK,
  description={Schramm and Koonin \cite{SCHRAMMKOONIN1991}}
}

\newglossaryentry{SPVHGLOSS}{
  name=SPVH,
  description={Salpeter and Van Horn \cite{SPVH1969}}
}

\newtheoremstyle{dtm}
  {0pt}
  {0pt}
  {\slshape}
  {0pt}
  {\bfseries}
  {. }
  {0pt}
  {}
\theoremstyle{dtm}

\author{Joseph L. Hellmers}

\title{EFFECTS OF PYCNONUCLEAR REACTIONS \\
  ON \\
  ACCRETING WHITE DWARF \\
  STELLAR EVOLUTION}
\titleheading{Effects of Pycnonuclear Reactions
  on \\ Accreting White Dwarf \\
  Stellar Evolution}

\degreeONE{Master of Science}
\degreeTWO{in}
\degreeTHREE{Computational Sciences}

\gradyear{2019}
\submitdate{Fall 2019}

\committeechair{Fridolin Weber}
\committeechairdept{Department of Physics}

\committeesecond{Calvin Johnson}
\committeeseconddept{Department of Physics}

\committeethird{William Welsh}
\committeethirddept{Department of Astronomy}

\begin{document}

\maketitle

\makesignature

\begin{copyrightpage}
  Copyright~\copyright~2019 \\
  by \\
  Joseph L. Hellmers
\end{copyrightpage}

%
\begin{dedication}
  \vspace{3in}
  \centering
To Ron and Carol Hellmers, who got me started on the right foot, \\
to Paula, who straightened me out quite a bit, \\
and to Lindsay, who supports and sustains me in all the best ways.
\end{dedication}


%
\begin{epigraph}
\noindent The heavens declare the glory of \\*
\indent God; \\*
the skies proclaim the work of His \\*
\indent hands. \\*
Day after day they pour forth speech; \\*
\indent night after night they reveal \\*
\indent\indent knowledge.
  \begin{flushright}
    -- Psalms 19:1-2
  \end{flushright}
\end{epigraph}

\begin{abstract}

This document presents a numerical study of the effects of high density nuclear reactions, pycnonuclear reactions, on the evolution of white dwarfs 
that accrete matter from a companion star.
Primarily we seek observable properties that might be different when we include these nuclear reactions with varying microphysical assumptions.
These different assumptions include effective nucleon-nucleon interactions, lattice structures and crystal polarization characteristics.
Our results indicate that, although we find some significant differences in pycnonuclear reaction rates, the differences do not result in observable 
differences in the white dwarf evolution.
Secondarily we examine the differences in observable properties when pycnonuclear reactions are assumed to happen and when they are assumed to not 
happen.
In this case, our calculations indicate there can be observable differences in the radius, $R$, and the effective temperature, $T_{eff}$, at the 
stopping condition we use for our simulations, $L_{nuc} > 10^{8}\thinspace \rm{L}_{\odot}$, that is the total rate of energy released by nuclear 
fusion is greater than $10^{8}$ solar luminosities.
Thirdly we explore the often neglected effect of temperature enhancement to pycnonuclear rates.

\end{abstract}

\tableofcontents

\listoftables

\listoffigures

\begin{glossarypage}
  \glsaddall\printglossary[title=]
\end{glossarypage}

\begin{acknowledgments}
My first taste of the study of high density nuclear reactions was the introduction Dr.Fridolin Weber gave me when he oversaw the work I did on my Bachelor of Science degree in
Physics at SDSU.
Subsequent to that we have continued to collaborate on this and other topics, and I expect we will continue to do so for quite awhile.
For this I thank him.
I also thank him for the guidance in the development of this thesis, both in the scientific subject matter and with the actual thesis process itself.
Most importantly, I have the deepest gratitude for his patience with this part-time student and researcher.

My journey to the completion of the requirements for a Master of Science degree has been a long and bumpy ride.  
It is impossible to imagine how it would have been possible without the help and long-suffering patience of Dr.Jos\`{e} Castillo.
Many thanks to him for everything he has done for me.

As a member of my thesis committee, Dr.William Welsh has given me helpful pointers, both with to respect to the stability of white dwarfs and about the content and format of my 
thesis.
I thank him for his involvement.

When I first contemplated returning to SDSU to complete my B.S. in Physics the first person I talked to was Dr.Calvin Johnson.
I do not know if he remembers this, but when I expressed my desire to perform research he, let me know I had many years to go before I could contribute anything.
This was in 2004, and 15 years later I am completing this thesis, so I guess he was right!
Regardless, I highly appreciate his participation as a member of my thesis committee.

The analysis performed for this thesis required the generation of more than 24 models of accreting white dwarf evolution.
Sometimes, after making program modifications, I needed to rerun the complete set.
I could have done this using my own personal computer, but it would have been very time consuming and prone to error.
Dr.Mary Thomas allowed me to have access to allocations on the San Diego Supercomputer Center's Comet system.
This enabled me to run the models in a very controlled environment. and allowed me to use my personal computer for other activities while those models ran.
Many thanks to her, and to the Comet and XSEDE support staff.

During the production of this thesis, my main source of employment has been with Dassault Syst\'emes.
I thank this company for their generous education reimbursement policy.
More specifically I thank my supervisors, Mike Peeler, Ton Van Dalen, and especially Christian Berger, for their patience with my class schedule and approval of the 
reimbursements for my course work.

Finally, I thank my wife Lindsay.
Besides happily proof reading this thesis, she has been an unending source of inspiration and support.
She has done a wonderful job taking care of me and our family during my educational endeavors, and I am deeply appreciative of her and all she has done and has endured.

\end{acknowledgments}

%
%

%
%
%

\chapter{INTRODUCTION}
\label{chap:intro}

The main purpose of this research is to determine how high density nuclear reactions, pycnonuclear reactions, might affect the observable 
properties of a white dwarf (WD) accreting material from a close companion star.
White dwarfs (WDs) are the remnants of stars similar in mass to our sun, the masses of which are supported by electron degeneracy pressure against further 
stellar collapse.
Normally an isolated WD would simply radiate away all of its heat on a timescale of billions of years.
When WDs are in a tight enough orbit around a main sequence (MS) star, however, it is possible for mass to transfer from the MS star through a process called 
accretion, resulting in a certain amount of stellar evolutionary dynamism on the part of the WD, and to a lesser extent, on the further evolution of the MS star.
If the mass accretion rate is high enough and lasts long enough, various types of high energy events can occur, such as Dwarf Novae, 
Classical Novae, Cataclysmic Variables, and Type Ia Supernovae.

In our studies we will not be looking for a particular event, because in order to simulate such events we would need to use fully three-dimensional 
models with timescales that would be very small.
The tools we use here do not support such modeling.
Instead we will assume that the WD star has reached reached a point very near the Chandrasekhar Limit when the total power generated by nuclear 
reactions is greater than $\rm 10^8\thinspace L_{\odot}$.
This assumption is justified by observing Figure \ref{fig:evolveNucPowerTime} which shows that the power spikes very sharply, indicating that a detonation is occurring

When determining the effects of pycnonuclear reactions on these types of systems, we will primarily look at variations in observable properties of 
WDs once this criterion has been met.
We will also examine differences in some of the internal characteristics of the WD star based on the different micro-physical details of our 
models

In addition, we seek to determine how variations in the nuclear reaction rates based vary based on different micro physics assumptions.
This differences in micro physics are nucleon-nucleon interaction, the lattice structure unit cell, and different approximations for crystal lattice polarization effects.

We will also consider issues related to the crystallization of white dwarf matter, non-zero temperature pycnonuclear reaction enhancement, and 
whether pycnonuclear reactions might affect the proton to neutron ratio of the WD's material, and hence affect the overall stability of the configuration 
with respect to modifications of the standard Chandrasekhar limit.

In order for our simulations not to be disturbed by various bursts and flashes near the surface, we will assume that the accreted material contains 
no hydrogen or helium.
Instead we will assume accreted matter is a mixture of 25\% carbon and 75\% oxygen.
This composition of accreted matter typically would occur if the donor star were another white dwarf.
We will also assume that the only significant pycnonuclear reactions occur for $\rm ^{12}C$.

\section{Subject Areas}

This thesis combines subject matter from four areas.

\begin{description}[font=$\bullet$~\normalfont\scshape\color{red!50!black}]
\item [White Dwarfs] The end point evolution for stars that have mass $\lesssim 4 \rm{M}_\odot$
\item [Accretion Processes] The transfer of mass from main-sequence stars to white dwarf companion stars
\item [Pycnonuclear Reactions] Nuclear fusion reactions at very high densities
\item [Nucleon-Nucleon Interaction] The potential energy due to the strong nuclear force
\end{description}

In order to define terms, introduce concepts, and generally set the stage for the following chapters, we present the following very brief historical introduction
and overview of these subjects.

\subsection{White Dwarfs}

In 1915 Adams reported spectrographic observations of Sirius' companion star \cite{ADAMS1915}, Sirius B, which indicated that the temperature of the
companion was similar to the temperature of Sirius A, instead of similar to a red dwarf star as was expected from its low luminosity.
It was already known that the mass of Sirius B is approximately the same as that of the sun.
With a low luminosity and a high temperature, it was apparent that the radius of Sirius B must be smaller than the radius of the earth, indicating
a very high density, considered to be outlandishly high.

In 1924, using general relativity, Eddington \cite{EDDINGTON1924} predicted the gravitational red shift for such a massive yet small object.
The following year, Adams was able to observe a red shift in accordance with Eddington's prediction, confirming the high density nature of Sirius B 
and adding more credibility to Einstein's general relativity.

Given what was considered a physically unnatural density for these objects theorists sought to understand their micro-physics
Elaborating on ideas of Eddington, in 1926 Fowler demonstrated that Fermi-Dirac statistics applied to WD matter could result in values of 
pressure able to sustain the equilibrium of these stars without other sources of energy \cite{FOWLER1926}.
Continuing in this line of reasoning, Chandrasekhar calculated that there would be a limit to how massive a white dwarf (a term originally 
coined by Luyten and popularized by Eddington) could be \cite{CHANDRASEKHAR1931}.
Interestingly the maximum value of the mass he calculated at the time was 0.91 solar masses.  The discrepancy between this value and the normally 
quoted value today (1.457 solar masses) is due to the number assumed for the molecular weight.
The molecular weight was assumed in 1931 to be an average of 2.5, whereas the more modern value is 2.

\subsection{Accretion Processes}

Astrophysical gravitational mass accretion is a very rich and well developed theoretical subject.
The processes related to the accretion of matter in large gravitation gradient fields can be used to explain systems from planetary and stellar 
systems to active galactic nuclei.

Some of the initial ideas and calculations related to accretion and the resulting accretion disks were done by Kant, Laplace, and Legendre in the 18th century.
In the late 1940s and early 1950s, von Weizs{\"a}cker and von L{\"u}st developed the modern theory and mathematics of accretion processes \cite{PRINGLE1981}.

Starting in the mid-1960s, astrophysicists began to realize that many of the galactic X-ray sources being discovered seemed to be 
associated with binary systems where one of the companion stars is a white dwarf or neutron star \cite{PRENDERGAST1968}.
Recently Shara et al. \cite{SHARA2018} have determined that an average accretion rate for Cataclysmic Variables is 
$\rm 10^{-7}$ to $\rm 10^{-8}\thinspace M_{\odot}$ per year.
For our simulations we use an accretion mass rate of $\rm 10^{-9}\thinspace M_{\odot}$ per year.

\subsection{Nucleon-Nucleon Interactions}

We will probe the effects of different formulations of the nucleon-nucleon interaction on calculated pycnonuclear reaction rates, and examine how these variations may change the evolution of the 
accreting system.
The history of nucleon-nucleon interaction theory is far too involved to go into much detail here.
Instead we will only give a very high-level overview of this history, and a somewhat more detailed description of more recent research related to the effective nucleon-nucleon interactions we use for 
this study.

Early in the history of nuclear physics it was determined that there is a ``mass defect'' in the nuclei of the atoms of the lighter elements.
For example, the mass of $\rm{He^4_2}$ is less than that of the constituent two neutrons and two protons.
Wigner \cite{WIGNER1933}  demonstrated that if we assume the potential energy between protons and neutrons is in the shape of a well, an attractive force, it is possible to explain and calculate the 
mass defects of the both deuterium and hydrogen nuclei.
In 1935, motivated by the quantum field theory of electrodynamics description of the photon as an exchange particle for that force, Yukawa \cite{YUKAWA1935} developed an initial idea about a heavy 
exchange particle for the nuclear force.
He also suggested an exchange particle for the weak nuclear force in the 1935 paper as well.

When two carbon-12 nuclei collide, we know that this is not a two-body problem, especially if the collision is inelastic.
In the simplest scenario this is a quantum 24 body problem, plus certain number of mesons.
For a completely \emph{ab initio} treatment we would have to consider quarks and gluons, which would amount to $3\times 24=72$ quarks as well as an indeterminate number of gluons.
These types of quantum many-body systems are very difficult calculate.
One way of dealing with interacting nuclei which sidesteps the many-body nature of this physical system is to use a so called Optical 
Potential.
In 1958 Feshbach \cite{FESHBACH1958} reasoned that a particle entering the vicinity of a nucleus could be thought of as a light wave 
traveling through some sort of media.

In 1979 Satchler and Love \cite{SatchlerLove1979} codified a nucleon-nucleon interaction based on earlier work by Bertsch et al. \cite{BERTSCH1977}.
This form of the interaction is called M3Y, due to there being a sum of three Yukawa terms.
The three terms include a one pion exchange potential for long-range interactions, greater than 0.4 fm, a multi-pion exchange component for a range 
$\sim$ 0.4 fm and a term based on a form designed to mimic the G-matrix elements of the Reid soft-core potential.

Another nucleon-nucleon potential that interests us here is the S\~{a}o Paulo potential.
This potential was originally derived by Chamon et al. in 1997, see \cite{CHAMON1997} and references therein.
They expanded on Jackson and Johnson's \cite{JACKSON1974249} single folding formulation to explain alpha scattering with nuclei data, which enabled them to extend Perey-Buck non-local potential effects to heavier ions.
Predecessor theses have used the nucleon-nucleon potential \cite{GOLFTHESIS,RYANTHESIS}, and the default pycnonuclear reaction rates of MESA use it as well.

A third nucleon-nucleon interaction for which we will perform pycnonuclear reaction rate calculations is based on Relativistic Mean Field (RMF) theory.
In 1974 Walecka used RMF theory to formulate the nucleon-nucleon interaction at very high densities \cite{WALECKA1974}.
Very generally, the approach is to use isoscalar scalar meson ($\sigma$-meson) exchange between nucleons for the attractive component of the interaction; the repulsive component is attributed to 
the exchange of heavier isoscalar vector mesons ($\omega$-mesons).  
For one-boson-exchange (OBE) subsequent research has shown that three additional mesons need to be considered to match experimental data: the $\delta$, $\rho$, and $\pi$ mesons.
In 2005 Serra et al. \cite{SERRA2005} used RMF principles to derive and expressions for the nucleon-nucleon interaction, taking into account density dependence, many-body interactions, and 
multi-scattering effects.
Singh et al. \cite{SINGH2012,SINGH2012C} found an optical potential expression for the nucleon-nucleon interaction in 2012, and compared it favorably to the M3Y expression.
We use the parameterizations from this 2012 article in our calculations of pycnonuclear reaction rates for RMF-based nucleon-nucleon effective potentials.

\subsection{Pycnonuclear Reaction Rates}

The term we use for the high density reactions we are studying in this thesis was coined by Cameron in 1959 \cite{CAMERON1959}.
The prefix ``pycno'' is an Anglicization of the Greek word \emph{pyknos}, meaning compact, dense, or thick.

Early attempts in the 1940s and 1950s to calculate high density nuclear reaction rates were based on one-dimensional potentials \cite{WILDHACK1940}, and assumed electron screening with a uniform
electron charge distribution \cite{CAMERON1959}.
In 1969 the seminal paper for pycnonuclear reactions was published by Salpeter and Van Horn \cite{SPVH1969}.
That paper included advanced treatments for electron screening, included the capability to calculate rates for mixed nuclei reactions, i.e. carbon-oxygen reactions as well as carbon-carbon reactions.
Also, that paper included formulae for temperature-enhanced reactions rates.
Reaction rates were given for both static and relaxed lattice polarization, and assumed a body-centered (bcc) lattice structure.

In 1991, Schramm and Koonin examined the effects of different approximations for lattice polarization, for which they found little effect \cite{SCHRAMMKOONIN1991}.
What they did provide which is useful for us, however, are rates when a face-centered (fcc) lattice structure is assumed.
The reaction rates provided were only applicable to carbon-carbon reactions.

Gasques et al. \cite{G05} developed a set of analytical formulae for calculating reaction rates for densities firmly in the pycnonuclear reaction regime and included temperature enhancement in density
and temperature regimes where that enhancement might occur.
As stated previously, they used the S\~{a}o Paulo nucleon-nucleon interaction.
Both fcc and bcc crystal structures are taken into account in these regimes and multi-component reactions rates can be calculated.

\section{Overview of Our Approach}

Very roughly, our approach is generate pycnonuclear reaction rates for a variety of cases, and then use a stellar evolution modeling tool, MESA, to determine what differences exist in the 
physical characteristics of an accreting white dwarfs based on the different way of calculating those pycnonuclear reaction rates.
The evolution simulations will be carried out until temperature  or total nuclear energy generation thresholds are reached.
For the final model we will observe differences in properties such as age, luminosity, mass, radius, temperature, total nuclear energy, and proton to neutron ratio.
We will also examine how the properties change during the accretion process for the different models we generate.

\chapter{MESA}

MESA is a scientific software application that generates models for an impressive variety of stellar systems and other systems which are massive enough to maintain a spherical shape such as 
gas giant planets \cite{MESA2011,MESA2013,MESA2015,MESA2018,MESA2019}.
There are several hundred parameters that specify various characteristics of the model.
In this chapter we review MESA's capabilities as directly related to white dwarf evolution.

MESA computes properties for a set of shells, or zones, that have the same basic conditions.
The size and location (radius) of these zones will change as the evolutionary model progresses.
Profiles are generated which contain the important stellar properties of the white dwarf as a function of radius.
As these properties change at slower or faster rates MESA adjusts the time scale between these profiles.
The time difference between profiles can be millions of years or fractions of a second depending on how dynamic the processes in a star occur.

MESA has the capability to have core hydrodynamic solvers and nuclear reaction networks use \texttt{OpenMP}.
\texttt{OpenMP} is method to parallelize code in a shared memory computational environment.
Typically this means when the processes are all running on the same node.
Almost all of our models were run with two \texttt{OpenMP} threads, the exception being the case when we attempted to include temperature enhanced pycnonuclear reactions.

\section{White Dwarf Modeling}

MESA handles a variety of equations of state based on the state, temperature, and density in the zones being computed for a particular time step
For white dwarfs the main equations of state that are utilized are referred to as SCVH (from Saumon et al.) and HELM (from Timmes and Swesty) \cite{MESA2019}.
In our models the more important equation of state is the HELM equation of state because it operates at higher densities, $\rm \rho > 10^3 \thinspace g/cm^3$ and, and temperatures below $10^8$ K.
This equation of state is a table of data interpolating of the Helmholtz free energy that uses a biquintic Hermite polynomial as the interpolating function \cite{TIMMESSWESTY2000}.

One of the characteristics of stellar evolution that is important to model correctly is diffusion of nuclei species within the matter of the star.
Typically this diffusion is calculated using Burgers equations.
In order to determine diffusion characteristics of degenerate and strongly coupled matter found in white dwarfs and neutron stars, a modified formalism of the cross-section calculation needs to be 
used \cite{MESA2018,MESA2015}.
MESA has extensions that result in the proper sedimentation of the nuclei in the interior of the white dwarf.

Besides the energy released by nuclear reactions, MESA includes the ability to model both the latent heat released by crystallization and by the contraction of material due to gravity.
MESA assumes that crystallization in the white dwarf occurs for the plasma coupling parameter, $\Gamma$, in the range of 150 to 175 \cite{MESA2018}.
This range may be tighten if desired.

It is possible to model shock propagation in white dwarfs with MESA.  It is also possible to model the evolution of both stars in a compact binary that are exchanging mass.
We do neither of these here.

\section{Mass Accretion}
MESA has the ability to model accretion by specifying a constant accretion rate in solar masses per year, adding a Fortran routine that implements an accretion rate that varies in a developer 
specified way, or creating a wrapper program to vary the accretion rate for each time step
When specifying a constant accretion rate, one can also specify abundances of the accreting material.
The accretion of material is assumed to occur in the outermost zones.
In order to determine the zones that have mass added to them, MESA starts with the outermost zone and keep adding zones until the mass of all selected zones is approximately equal to the mass that is
being accreted.

When matter is accreting onto a white dwarf (or any other star) there will be a certain amount of compressional heating.
MESA models this effect as well.

MESA is also able to model the effects of mass loss due the Eddington limit being exceeded.  We do not include these effects for this work.

\chapter{COMPUTATIONAL PROCEDURE}

\begin{figure}[ht]
  \centering
  \begin{minipage}{4.5in}
    \includegraphics[width=\linewidth]{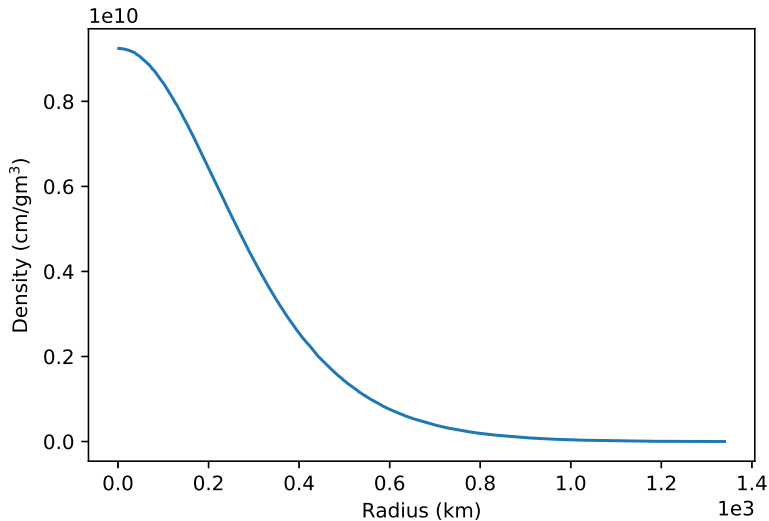}
    \caption{Final density profile of the G05 model.\label{fig:g05afinalrho}}
  \end{minipage}
\end{figure}

As can be seen from Equations~(\ref{eqn:invLenLambda}) and~(\ref{eqn:SPVHRate}), the pycnonuclear reaction rate is highly dependent on density of the white dwarf material.
More detailed calculations show that between the densities of $\rm \rho\thinspace=10^9$ to $\rm 10^{11}$ $\rm g/cm^3$ the pycnonuclear reaction rate increases from effectively 0 to approximately $10^{35}$
reactions for cubic cm per second.
Since computing these reactions rates can be computationally costly, and we wish to look at the results of reaction rates with a variety of assumptions, we will only calculate them for the
range of densities that are needed in the astrophysical system we model here.
To determine the density range we need to calculate, we examine the density profile of the final model for the MESA-provided model, which assumes the G05 reaction rates.
These results shown in Figure~\ref{fig:g05afinalrho}, show the maximum density is slightly less than $\rm 10^{10}\thinspace g/cm^3$.
In order to make sure we can cover the complete range of densities, we will actually calculate the rate up to $\rm 10^{11}\thinspace g/cm^3$. 

\begin{figure}[ht]
  \centering
  \begin{minipage}{4.5in}
    \includegraphics[width=\linewidth]{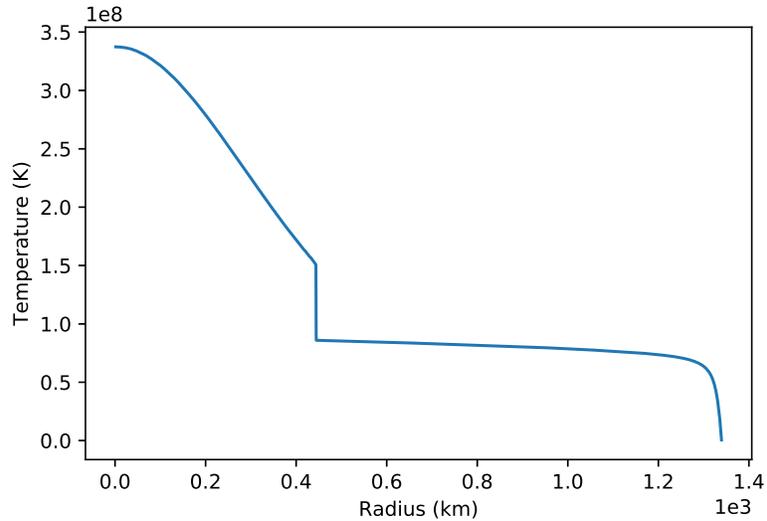}
    \caption{Final temperature profile of the G05 model.\label{fig:g05tempfinal}}
  \end{minipage}
\end{figure}

In this study we will also be evaluating how temperature enhancement of pycnonuclear rates will affect white dwarf evolution.
Since this is the case we also use the final temperature profile of the G05 model to analyze whether temperature enhancement is important, see Figure~\ref{fig:g05tempfinal}.

\begin{figure}[ht]
  \centering
  \begin{minipage}{4.5in}
    \includegraphics[width=\linewidth]{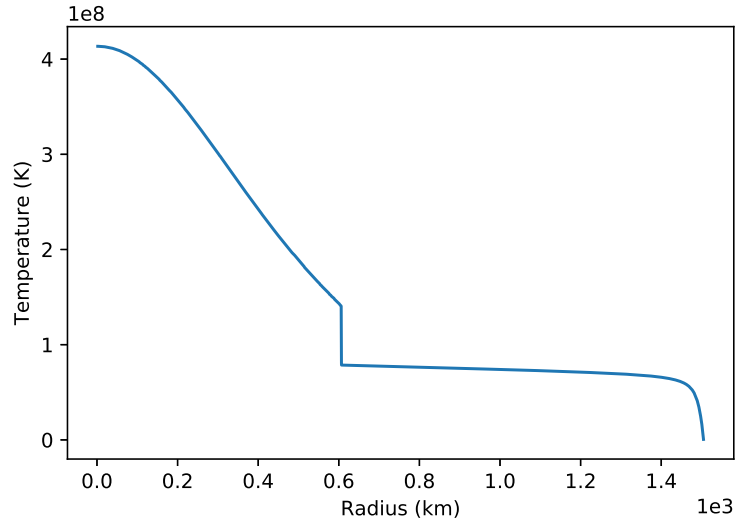}
    \caption{Final temperature profile of the accreting white dwarf model assuming no pycnonuclear reactions.\label{fig:nopcynotempfinal}}
  \end{minipage}
\end{figure}

It is interesting to note that we have a discontinuity in the temperature of this model at a distance of a little over 400 km from the center of the star.
Part of our analysis will be to attempt to determine what is happening at this region of the star to cause this feature.
We note this discontinuity does not seem to be related to the existence pycnonuclear reaction or lack thereof by observing that the temperature profile the final model of an accreting white dwarf 
model assuming no pycnonuclear reactions has the same feature at approximately 600 km.
See Figure~\ref{fig:nopcynotempfinal}.

\begin{table}[hbt]
  \centering
  \begin{minipage}{5.5in}
    \centering
    \caption{Matrix of Pycnonuclear Reaction Microphysics Assumptions\label{tab:microAssumptions}.}
    \begin{tabular}{|c|c|c|c|c|}    \hline
      Method & NN-Pot. & Cell & Approx & Model Name \\ \hline \hline
		G05 & S\~{a}o Paulo & bcc & Static & wd\textunderscore ignite\textunderscore g05 \\ \hline
		SPVH & M3Y & bcc & Static & wd\textunderscore ignite\textunderscore m3y\textunderscore spvh \\ \hline
		SK & M3Y & bcc & Relaxed & wd\textunderscore ignite\textunderscore m3y\textunderscore bcc\textunderscore relaxed \\ \hline
		SK & M3Y & bcc & Static & wd\textunderscore ignite\textunderscore m3y\textunderscore bcc\textunderscore static \\ \hline
		SK & M3Y & bcc & WS  & wd\textunderscore ignite\textunderscore m3y\textunderscore bcc\textunderscore ws \\ \hline
		SK & M3Y & fcc & Relaxed & wd\textunderscore ignite\textunderscore m3y\textunderscore fcc\textunderscore relaxed  \\ \hline
		SK & M3Y & fcc & Static & wd\textunderscore ignite\textunderscore m3y\textunderscore fcc\textunderscore static  \\ \hline
		SK & M3Y & fcc & WS & wd\textunderscore ignite\textunderscore m3y\textunderscore fcc\textunderscore ws \\ \hline
		SPVH & S\~{a}o Paulo & bcc & Static & wd\textunderscore ignite\textunderscore sp\textunderscore spvh \\ \hline
		SK & S\~{a}o Paulo & bcc & Relaxed & wd\textunderscore ignite\textunderscore sp\textunderscore bcc\textunderscore relaxed \\ \hline
		SK & S\~{a}o Paulo & bcc & Static & wd\textunderscore ignite\textunderscore sp\textunderscore bcc\textunderscore static   \\ \hline
		SK & S\~{a}o Paulo & bcc & WS & wd\textunderscore ignite\textunderscore sp\textunderscore bcc\textunderscore ws   \\ \hline
		SK & S\~{a}o Paulo & fcc & Relaxed & wd\textunderscore ignite\textunderscore sp\textunderscore fcc\textunderscore relaxed \\ \hline
		SK & S\~{a}o Paulo & fcc & Static & wd\textunderscore ignite\textunderscore sp\textunderscore fcc\textunderscore static   \\ \hline
		SK & S\~{a}o Paulo & fcc & WS & wd\textunderscore ignite\textunderscore sp\textunderscore fcc\textunderscore ws   \\ \hline
		SPVH & RMF & bcc & Static  & wd\textunderscore ignite\textunderscore rmf\textunderscore spvh \\ \hline
		SK & RMF & bcc & Relaxed & wd\textunderscore ignite\textunderscore rmf\textunderscore bcc\textunderscore relaxed \\ \hline
		SK & RMF & bcc & Static & wd\textunderscore ignite\textunderscore rmf\textunderscore bcc\textunderscore static   \\ \hline
		SK & RMF & bcc & WS & wd\textunderscore ignite\textunderscore rmf\textunderscore bcc\textunderscore ws   \\ \hline
		SK & RMF & fcc & Relaxed & wd\textunderscore ignite\textunderscore rmf\textunderscore fcc\textunderscore relaxed \\ \hline
		SK & RMF & fcc & Static & wd\textunderscore ignite\textunderscore rmf\textunderscore fcc\textunderscore static   \\ \hline
		SK & RMF & fcc & WS & wd\textunderscore ignite\textunderscore rmf\textunderscore fcc\textunderscore ws   \\ \hline
	\end{tabular}
  \end{minipage}
\end{table}

There are four basic factors we will be varying for the calculation of pycnonuclear reaction rates.
The first of these is the method of calculation.  
One method of calculation will be the one described in G05, which is already provided by MESA.  
The second method of calculation is the one used in the seminal work of SPVH.
The third method is the one developed for carbon-carbon reactions in SK.

The second factor we will vary in our pycnonuclear rate calculations is the nucleon-nucleon interaction.
In summary these are the M3Y potential, the S\~{a}o Paulo potential, and an effective potential from relativistic mean field theory.
These nucleon-nucleon interactions are discussed more thoroughly in Chapter~\ref{chapter:nnInteractions}.

The third micro-physical characteristic we will vary in our calculations is the assumed lattice structures of body-centered (bcc) and face-centered (fcc).

The final attribute we will vary in our pycnonuclear reaction rates is how the lattice structure responds to the interacting nuclei.
The possibilities include static, relaxed, and Wigner-Seitz (WS).
The static case assumes that the nuclei do not move during the interaction.
In the relaxed case, the nuclei move with no time delay during the interaction.
The WS case is similar to the relaxed case but is approximated by subtracting the energies of the WS spheres from the fused state of the nuclei \cite{G05}.

To use our calculated pycnonuclear reaction rates to generate accreting white dwarf models, we modify the appropriate program unit in MESA, \texttt{pycno.f90}, to use a cubic spline.
This cubic spline is calculated outside of the model and incorporated into \texttt{pycno.f90}.

In order to generate our accreting white dwarf models for the cases listed in Table~\ref{tab:microAssumptions} consistently and to allow for reproducibility we use the Comet computing system at the
San Diego Supercomputing Center.
By submitting jobs using a common scripting system and saving the data in a consistent manner, we have made a workflow that allows us to regenerate the models as needed and transfer all the data in an
easy to use fashion.
This allows us to make changes to the code and rerun everything again as needed.

Details of the programming changes and other computational matters are provided in Chapter~\ref{chapter:mesaMods} and Appendix \ref{appendix:compDetails}.

\chapter{NUCLEON-NUCLEON INTERACTION}
\label{chapter:nnInteractions}
In order to truly do an \emph{ab initio} determination of the nucleon-nucleon interaction one would need to use quark and gluon (partons) physics.
Quarks are constituents of the protons and neutrons that make up a nucleus and gluons are the boson exchange particles of the strong nuclear force between them.
While this is not a tractable problem yet, lattice Quantum Chromodynamics (QCD) is perhaps getting close.  
Here we will content ourselves with using effective nucleon-nucleon potential expressions cast in the form a central potential.

There are several characteristics that any nucleon-nucleon effective potential must adhere to in order to be consistent with experimental data.
See \cite{BertulaniNuc,KraneNuc}.
Proton-proton and neutron-neutron scattering results are almost equivalent after correcting for electrostatic repulsion of the protons.  
This necessitates that the nucleon-nucleon interaction has an approximate property called isospin independence, meaning the potential is almost the
same regardless of the whether the nuclear material is protonic or neutronic in nature.  

Another characteristic of the nucleon-nucleon interaction is that it is stronger than electrostatic repulsion at short distances, otherwise nuclear fusion would not be possible.  

Nucleon scattering experiments yield results that show the central density of nuclei remain roughly constant as the mass number increases.  
This indicates there should also be a repulsive core.
The M3Y and the RMF nucleon-nucleon interactions have this property but the S\~{a}o Paulo does not.

One interaction we will use is the so-called M3Y interaction given by Equation~(\ref{eqn:m3ypot}).
This interaction ws first described as such by Satchler and Love, \cite{SatchlerLove1979}, based on work by Bertsch et al. \cite{BERTSCH1977}.
The potential was derived from the Reid potential using the Bruekner G-Matrix theory.
The first term is an expression for the one pion exchange potential, and the second term is a potential that simulates multiple pion exchange.
This expression for the nucleon-nucleon interaction, as well as the expressions for the S\~{a}o Paulo and RMF, gives in units of MeV.

\begin{equation}
   V_{\mathit{NN},\mathit{m3y}} = 7999\frac{e^{-4r}}{4r}-2134\frac{e^{-2.5r}}{2.5r} - 262\delta (r) \label{eqn:m3ypot}
\end{equation}

The S\~{a}o Paulo potential given in Equation~(\ref{eqn:sppot}) was first formulated by Chamon et al. in 1997 \cite{CHAMON1997}.
In their work they assumed a double-folding model for the nuclear potential, and that non-locality could be used to remove density dependence.
One of the motivations for examining this effective potential is the fact that it is used for the pycnonuclear reaction rate calculations of Gasques et al. \cite{G05}.
In this formula $a_m$ is the diffusivity of the nuclei which is taken to be 0.3 fm in our calculations.

\begin{equation}
  V_{\mathit{NN},\mathit{sp}} = \frac{V_0}{64\pi a^3_m}e^{\frac{-r}{a_m}}(1+\frac{r}{a_m}+\frac{r^2}{3a^2_m}) \label{eqn:sppot}
\end{equation}

Another effective nucleon-nucleon interaction given by Equation~(\ref{eqn:rmfpot}) can be derived from relativistic mean field theory \cite{SINGH2012}. 
In this expression the three terms correspond to the exchange potentials for the $\omega$, $\rho$, and $\sigma$ mesons respectively.
Each term is proportional to the square of coupling constants, and has an exponential decrease as the mass of the appropriate meson increases.
Note that the mass needs to be divided by 197.3 MeV$\cdot$fm to get the mass in units of $\rm{fm}^{-1}$.

\begin{equation}
  V_{\mathit{NN},\mathit{rmf}} = \frac{g^2_\omega}{4\pi}\frac{e^{-m_\omega r}}{r}+\frac{g^2_\rho}{4\pi}\frac{e^{-m_\rho r}}{r}-\frac{g^2_\sigma}{4\pi}\frac{e^{-m_\sigma r}}{r}\label{eqn:rmfpot}
\end{equation}

The exact values for the coupling constants and meson masses are model dependent and given in Table~\ref{tab:rmfmodels}.

\begin{table}[hbt]
  \centering
  \begin{minipage}{6.5in}
    \centering
    \caption{Masses and Coupling Constants for Different RMF Formulations\label{tab:rmfmodels}.}
    \begin{tabular}{|c||c|c|c|c|c|c|}    \hline
      Formulation &	$m_\sigma$ MeV  &  $m_\omega$ (MeV) & $m_\rho$ (MeV) & $g_\sigma$ & $g_\omega$ & $g_\rho$ \\ \hline \hline
      HS & 520 & 783  & 770 & 10.47 & 13.80 & 8.08 \\ \hline
      Z & 551.31 & 780  & 763 & 11.19 & 13.83 & 10.89 \\ \hline
      W & 550 & 783  & - & 9.57 & 11.67 & - \\ \hline
      L1 & 550 & 783  & - & 10.30 & 12.60 & - \\ \hline
    \end{tabular}
  \end{minipage}
\end{table}

\begin{figure}[ht]
  \centering
  \begin{minipage}{4.5in}
    \includegraphics[width=\linewidth]{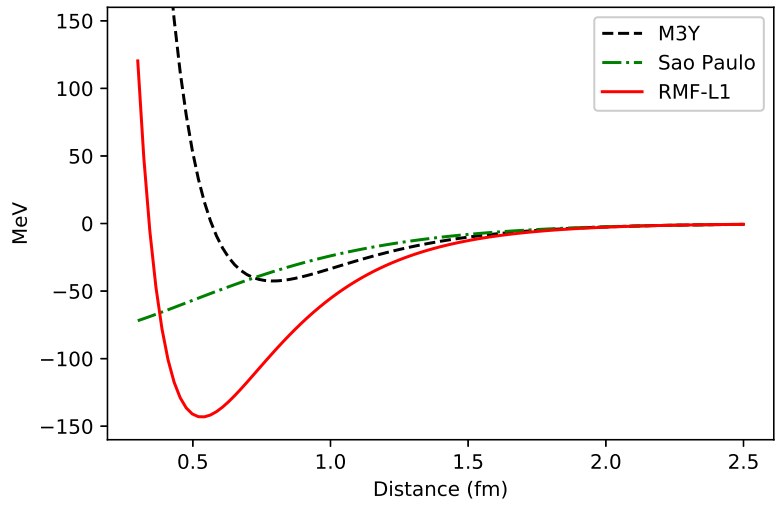}
    \caption{Comparison of nucleon-nucleon potentials.\label{fig:nnpotcompare}}
  \end{minipage}
\end{figure}

We have no basis for picking one model over the others.
One of our goals is to see if we can make predictions of observable properties of accreting white dwarf models that might be different for different nucleon-nucleon reactions.
In Singh et al. \cite{SINGH2012C} a graph is given the shows the results of the various RMF models against the M3Y interaction.
The L1 RMF model has the deepest energy well, far deeper than M3Y and the RMF models HS and Z, and a bit deeper than RMF model W.
We surmise that this will have the biggest positive effect on pycnonuclear reaction rates, hence we will use the L1 RMF model when calculating our RMF pycnonuclear reaction rates.
We present Figure~\ref{fig:nnpotcompare} for a visual representation of the differences in the nucleon-nucleon interactions as a function of distance between nuclei.

\chapter{PYCNONUCLEAR REACTION PHYSICS}

\begin{figure}[!htb]
  \centering
  \begin{minipage}{4.5in}
    \includegraphics[width=\linewidth]{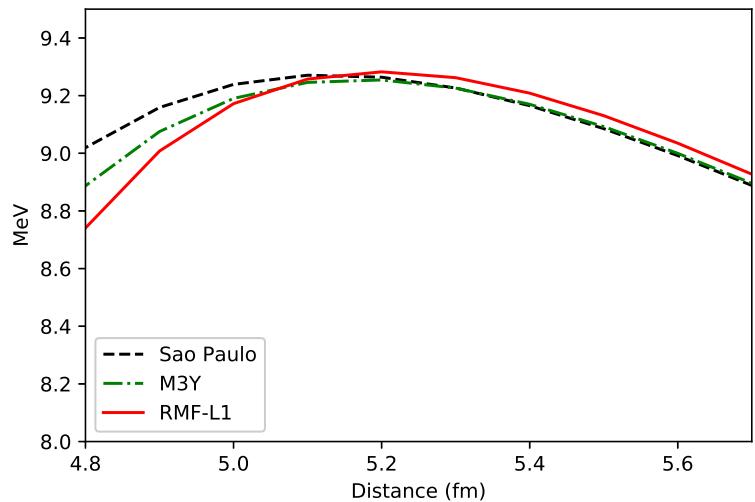}
    \caption{This plot shows the potential barrier that a nuclei approaching from the right needs to tunnel through in order to fuse with the target nuclei.\label{fig:VTotPlotZoom}}
  \end{minipage}
\end{figure}

In order for nuclear fusion reactions to occur there must be a strong nuclear force between hadronic particles (i.e. particles composed of quarks) that is able to counteract the electrostatic repulsion
of completely ionized atoms when those atoms get close enough.
In the normal case of main sequence stars, the nuclear fusion reactions are thermal in nature.
When the gaseous plasma of these stars is hot enough and dense enough, nuclei will collide with a high relative speed.
This allows nuclei near enough to quantum mechanically tunnel through the energy barrier and then fuse.
See Figure~\ref{fig:VTotPlotZoom} for an example of the energy that must be tunneled through in order for fusion reactions to occur.

Pycnonuclear reactions are thought to occur in white dwarfs and neutron stars because the nuclei are in close proximity due to the gravitational collapse of those objects.
In these objects the density is so high that the spacing between nuclei is $\sim$ 10 to 100 fm.
At these densities the probability of nuclei tunneling through their respective potential barriers becomes appreciable, and it seems likely that the nuclear fusion reactions will start occurring with
enough frequency to start to affect the energetics of the stellar object.

In the first section of this chapter, we will discuss the basic properties we need to determine in order to calculate the pycnonuclear reaction rates including the density as a function of distance 
from the center of fusing nuclei, the nuclear radius expression, and the reduced mass.  In order to calculate the total effective nuclear potential between two nuclei we need to calculate the so called double
folding potential, which will be introduced in section 5.2.
Reaction rates between particles are calculated by using a reaction cross-section.
Pycnonuclear reaction rates, however, are very low energy, that is, the relative velocity between the nuclei is very low.
This motivates us to use a parameter called the astrophysical S-factor which is inversely proportional to the reaction cross-section; we discuss how this is calculated in section 5.3.
In section 5.4 we describe how to combine the calculations discussed previously to compute pycnonuclear reaction rates for a multi-component in high density matter as originally derived in SPVH.
Schramm and Koonin performed some refinement of the method of SPVH for carbon-carbon nuclei fusion.
We present that calculation in section 5.5.
G05 created a multi-regime calculation for carbon-carbon reaction rates.
In section 5.6 we discuss these calculations as they pertain to pycnonuclear reactions.
Triple-$\alpha$ nuclear reactions can proceed for high densities, and have the possibility of being quite energetic.
In section 5.7 we briefly discuss these, mostly to discount them in the case of accreting white dwarf systems that we model here.

\section{Basic Nuclear Parameters}

In order to calculate the reaction rates we need a few basic nuclear properties.
Particularly, to perform calculations of the double folding potential, $V_{fold}$, we need to know the nuclear radius of the nuclei involved as well as nuclear density.
We assume the nuclear density to be a two parameter Fermi distribution (2pF).
Note that this nuclear density is a number density.
In the expression for $\rho(r)$, Equation~(\ref{eqn:nucleardensity}), $a$ is the diffusivity parameter, taken to be 0.5 fm, and $R_0$ is the nuclear radius in fm.

\begin{equation}
  \rho(r) = \frac{\rho_0}{1+e^{\frac{r-R_0}{a}}}\label{eqn:nucleardensity}
\end{equation}

This 2pF distribution	 must be normalized to the atomic mass number of the nucleus.
In general neutron density and proton density of nuclei are not exactly coincident.
In this thesis we will consider only a total number density for nuclear matter, be it neutron or proton matter.

\begin{equation}
  4\pi\int^{\infty}_0 \rho(r)r^2dr=A
\end{equation}

Both the diffusivity parameter and the radius are determined by best fits to experiment and from other calculations.
In Chamon et. al \cite{Chamon2002} this relation is derived by fitting data with both experimental electron scattering and Dirac Hartree Bogoliubov (DHB) calculations.
The resulting expression yields nuclear radii with the units of femtometers, fermi.

Based upon a linear fit of the of $\rm{A^{1/3}}$ to the data and calculations the radius (in fm) has a form in Equation~(\ref{eqn:nuclearRadius}), \cite{CHAMON2003}.
\begin{equation}
  R_0 = 1.31A^{1/3}-0.84 \label{eqn:nuclearRadius}
\end{equation}

We will also need the reduced mass between the two ``colliding'' nuclei, given with full two component generality in Equation~\ref{eqn:reducedMassFull}.

\begin{equation}
  \mu = \frac{(Z_1m_p+N_1m_n)^2}{(Z_1+Z_2)m_p+(N_1+N_2)m_n} \label{eqn:reducedMassFull}
\end{equation}

\section{Nucleon-Nucleon Folding Potential}

\begin{figure}[!htb]
  \centering
  \begin{minipage}{4.5in}
    \includegraphics[width=\linewidth]{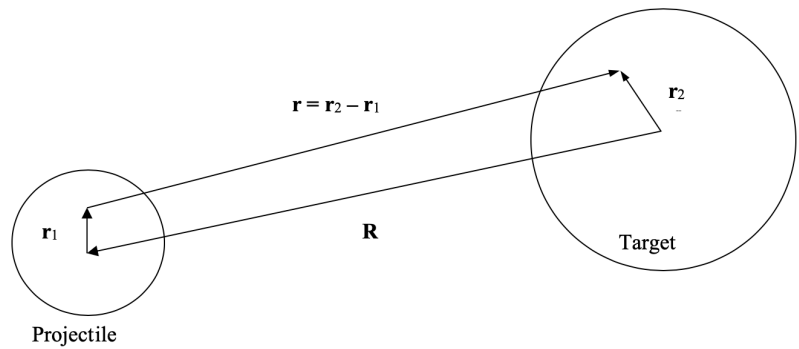}
    \caption{Diagram of nuclei for folding potential double triple integral.\label{fig:FoldingDiagram}}
  \end{minipage}
\end{figure}

The nucleon-nucleon interaction potentials given in Equations~(\ref{eqn:m3ypot}),~(\ref{eqn:sppot}), and~(\ref{eqn:rmfpot}) evaluate to a value in units of energy divided by unity squared.
The folding potential is calculated by summing all the strong nuclear force attractions for the infinitesimal volume elements of the two nuclei, see Figure~\ref{fig:FoldingDiagram}.

\begin{equation}
  V_{N} = V_{fold}=\int\int\rho_1(\mathbf{r}_1)\rho_2(\mathbf{r}_2)V_{NN}(\mathbf{R}-\mathbf{r}_2+\mathbf{r}_1)d\mathbf{r}_1d\mathbf{r}_2 \label{eqn:vfold}
\end{equation}

For this study we assume the nuclei possess spherical symmetry.
This being the case, we could use Fourier analysis to evaluate the integrals.
However, we want to be able to extend our computations to deformed nuclei.
Therefore we implement a brute-force algorithm that is not as efficient for the spherically symmetric case as other algorithms, but can be more easily modified for the non spherical case.
Also, the method is amenable to computational parallelization.

\begin{figure}[!htb]
  \centering
  \begin{minipage}{4.5in}
    \includegraphics[width=\linewidth]{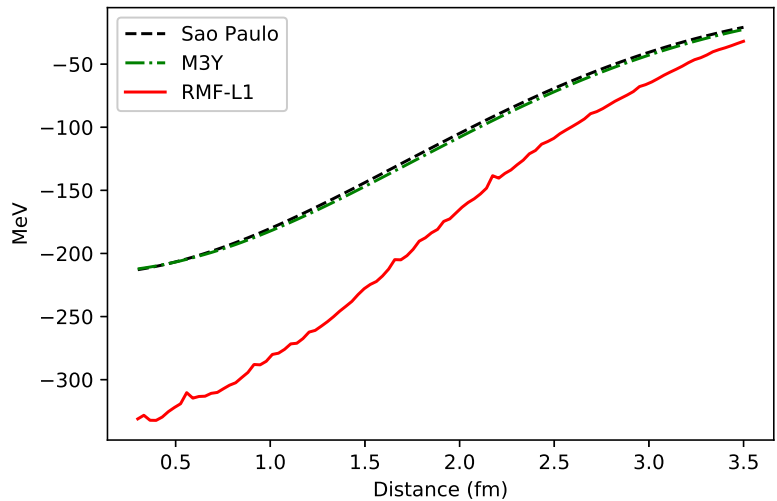}
    \caption{Folding potentials as a function of density.\label{fig:vfoldPlot}}
  \end{minipage}
\end{figure}

It can be seen in Figure~\ref{fig:vfoldPlot} that the folding potential for the M3Y and S\~{a}o Paulo nucleon-nucleon effective interaction  are practically identical and the RMF-L1 is quite a bit greater at closer distance.
That is, the potential well of the folding potential in the RMF-L1 nucleon-nucleon case is quite a bit deeper.

\begin{figure}[!htb]
  \centering
  \begin{minipage}{4.5in}
    \includegraphics[width=\linewidth]{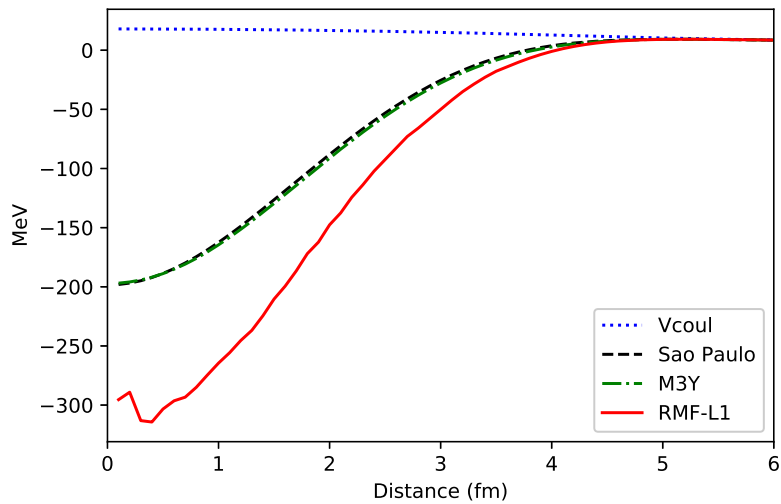}
    \caption{Total potentials, including electrostatic.\label{fig:VTotPlot}}
  \end{minipage}
\end{figure}

In order to calculate the astrophysical S-factor we simply need to compute the electrostatic potential and add it to the folding potential.
The results for this calculation are shown in Figure~\ref{fig:VTotPlot}.
To given an indication of the size of the nuclear force compared to the electric force, the electrostatic potential by itself is also shown.

\section{Astrophysical S-Factor}

The nuclear reactions we study here are intrinsically low energy from the perspective of the relative velocity of the ions \cite{ClaytonNucleosynthesis}.
It is worth noting here that it is not only pycnonuclear reactions that are low energy: even thermonuclear reactions typical in main sequence stars have relatively low energy compared to the 
reactions that are typically produced in earthly laboratories.
For low energy collisions, the cross-section is proportional to the geometric factor $\pi \lambda^2$ \cite{BHWD1983}, where $\lambda$ = h/p = h/mv, the de Broglie wavelength.

\begin{equation}
  \sigma(E) \propto \pi \lambda^2 \propto \frac{\pi h^2}{2mE} \propto \frac{1}{E}\label{eqn:crossSection}
\end{equation}

Here we are considering motion in a 2-body system with the origin at the center of mass.  The velocity is relative velocity in that system, and the mass is the reduced mass.
Equation~\ref{eqn:crossSection}, indicates that the interaction cross section rapidly goes to infinity as the energy becomes small.
Experimentally determined cross-sections for nuclei collisions bear this fact out, and extrapolating the cross section at low energies becomes problematic.

We are also able to examine the transmission coefficient of colliding nuclei.
This tunneling effect occurs because of the nature of the overall potential well between the interacting nuclei.
Thinking of the nuclear reaction as a barrier potential problem, due to the combined potential of the electrostatic repulsion and the nuclear interaction well, 
we recall the transmission coefficient or Gamow penetration factor, given in Equation~(\ref{eqn:GamowFactor}).
$v$ in this case is the relative velocity of the fusing nuclei.

\begin{equation}
  T = e^{-2\pi \eta},\medspace \eta = \frac{Z_1Z_2e^2}{hv}\label{eqn:GamowFactor}
\end{equation}

So we have good reasons to believe that the nuclei collisional cross section for the reaction is inversely proportional to the energy and directly proportional to the Gamow factor.
A function of energy, $S(E)$, is defined that allows us to write Equation~(\ref{eqn:crossSectionSFactor}).

\begin{equation}
\sigma(E)=\frac{S(E)}{E}e^{-2\pi \eta} \label{eqn:crossSectionSFactor}
\end{equation}

This is referred to as the astrophysical S-factor.
We can rewrite $S(E)$ as a function of energy, cross section and the Gamow factor, Equation~(\ref{eqn:SFactor}).
We shall see that the S-factor varies slowly with energy in the low energy regime.

\begin{equation}
S(E)=\sigma(E)E e^{-2\pi \eta} \label{eqn:SFactor}
\end{equation}

From Equation~(\ref{eqn:SFactor}) we can see that an expression for the cross section is required.
In order to compute the cross section, we need to be aware of the total potential that acts between the two colliding nuclei.
We have discussed in detail the nucleon-nucleon interaction, and we have mentioned the electrostatic potential, however it is important to understand that there could be a potential associated with 
centrifugal effects.

\begin{equation}
  V_{eff}(r,E,\ell)=V_C(r)+V_N(r,E)+\frac{\ell(\ell+1)\hbar^2}{2\mu r^2}
\end{equation}

The nuclear force potential is calculated from Equation~(\ref{eqn:vfold}).  We do this based on the various nucleon-nucleon potentials, $V_{NN}$, as defined in Equations~(\ref{eqn:m3ypot})
,~(\ref{eqn:sppot}), ~(\ref{eqn:rmfpot}).

The electrostatic term for the potential is the standard potential that is well known.
However, since we are seeking to perform calculations for quantum tunneling we need to include the potential inside the nuclei radius.
Hence we need to use an expression for the electrostatic repulsion that includes the regions inside the radii of the nuclei.
Also note that for the electrostatic potential we assume a uniform density of charge (i.e. protons) in the nucleus.
$R_1$ and $R_2$ are the radii for the nuclei that are fusing as given in Equation~(\ref{eqn:nuclearRadius}).
The electrostatic repulsion needs to include the fact that the proton nuclear material will overlap as the nuclei get closer together.

\begin{equation}
 V_C(r)=
  \begin{cases}
	(3(R_1 + R_2)^2-r^2)\frac{Z_1Z_2e^2}{2(R_1+R_2)^3},	& r < R_1 + R_2 \\
    \frac{Z_1Z_2e^2}{r},					& r \ge R_1 + R_2
  \end{cases} 
\end{equation}

For the centrifugal component of the total potential we note that $\ell$ corresponds to the $\ell$-wave in a partial-wave decomposition from basic quantum scattering theory.
In general $\ell$ is an integer that starts with 0 and can go to infinity.
In order for the colliding nuclei to actually fuse there must be a potential well, therefore the potentials that need to be calculated, in general, are for $\ell$ up to some
maximum, $\ell_{cr}$, where the potential well disappears.
In this thesis we consider only the $\ell=0$ wave, therefore the centrifugal term vanishes.

We can use an expression for fusion cross section based on the particle flux transmitted through the potential barrier \cite{G05}.
Here we terminate the sum at $\ell = 0$.

\begin{equation}
\sigma(E) = \frac{\pi \hbar^2}{2\mu E_0}\sum_{\ell=0}^{\ell_{cr}} (2\ell + 1)T_\ell = \frac{\pi \hbar^2}{2\mu E_0}T_0
\end{equation}

Using a heuristic method from Schiff \cite{schiff1968quantum}, we can find an expression for the transmission probability using the WKB approximation at the potential well turning points.

\begin{equation}
 V_{WKB} = \int^{r_2}_{r_1}\mathcal{F}(r)dr
\end{equation}

\begin{equation}
 \mathcal{F}(r)=
  \begin{cases}
	 + \sqrt{\frac{8\mu}{\hbar^2}[V_{eff}(r)-E_0]}dr,	&  V_{eff}  \ge E_0 \\
    - \int^{r_2}_{r_1}\sqrt{\frac{8\mu}{\hbar^2}[E_0-V_{eff}(r)]}dr	&  V_{eff} < E_0
  \end{cases} 
\end{equation}

\begin{equation}
   T_0 = \frac{1}{1+e^{V_{WKB}}}
\end{equation}

\begin{figure}[!htb]
  \centering
  \begin{minipage}{4.5in}
    \includegraphics[width=\linewidth]{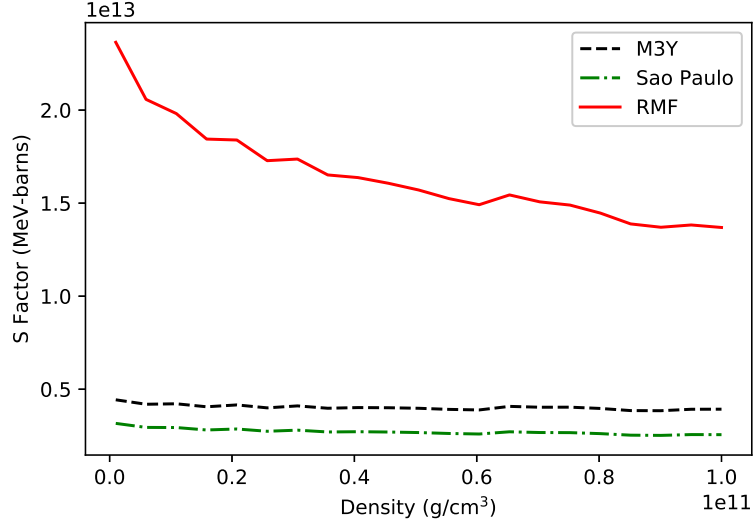}
    \caption{Astrophysical S-factor as a function of density for nucleon-nucleon interactions.\label{fig:SFactor}}
  \end{minipage}
\end{figure}

In order to finish this calculation we need the value for $E_0$.  We will go over this calculation in the section that discusses the pycnonuclear reaction rates as derived in SPVH.
For now we present the calculations for the three different nucleon-nucleon interactions we use in our rate calculations in Figure~\ref{fig:SFactor}.
As can be seen, the S-factor is varies only slowly for the RMF interaction, and stays essentially flat for the M3Y and S\~{a}o Paulo interactions.

\section{Salpeter-Van Horn Rates from 1969}

SPVH presented pycnonuclear reaction rate calculations taking into account many micro-physics effects.  One effect was to account for both weak and strong screening for a multi-component 
nuclear reaction.  Salpeter and Van Horn were able to develop interpolation methods to cover a wide range of atomic number nuclei and to be accurate for densities in the pycnonuclear regime.
In addition they developed methods for calculating the lattice and vibrational energies of the nuclei at the lattice points, and were able to include temperature enhancement of reaction rates.
In this chapter we only consider zero temperature reaction rates.
We will discuss temperature enhancement in Chapter \ref{chapter:tempAndCrystal}.

In order to calculate the background energy in which our pycnonuclear reaction occurs, $E_0$, we add the lattice energy and the vibrational energy of the ions at the lattice points.

\begin{equation}
 E_0 = E_{lattice} + E_{vib}
\end{equation}

There are several parameters the SPVH define in order to simplify computations.
One of these is the dimensionless inverse length parameter $\lambda$.
In this expression $\mu_e$ is the mean molecular weight per electron. 
The $X_i$ value in the expression for $\mu_e$, Equation~(\ref{eqn:electronmolwt}), is the percentage of the $i^{th}$ nuclear species.

\begin{equation}
  \lambda = \frac{A_1+A_2}{2A_{1}A_{2}Z_{1}Z_{2}}\left(\frac{1}{Z_1\mu_e}\frac{\rho}{1.3574 \times 10^{11} g\cdot cm^{-3}}\right)^{1/3} 
\label{eqn:invLenLambda}
\end{equation}

\begin{equation}
  \frac{1}{\mu_e} = \sum_{i=1}^{N}\left( \frac{\frac{X_i}{A_i}}{1+\frac{Z_i}{A_i}\frac{m_e}{\rm{amu}}} \right) \label{eqn:electronmolwt}
\end{equation}

Another defined parameter, motivated by the expression for the Bohr radius, is the characteristic radius $r^*$.

\begin{equation}
 r^* = \frac{A_1+A_2}{2A_1A_2Z_1Z_2}\frac{\hbar^2}{\rm{amu} \cdot e^2}
\end{equation}

From the characteristic radius we can then define a characteristic energy $E^*$.

\begin{equation}
 E^* = \frac{Z_1Z_2e^2}{r^*}
\end{equation}

Based on SPVH for bcc lattice structures, the lattice energy is given by Equation~(\ref{eqn:ELattice}). and the vibration energy is given by Equation~(\ref{eqn:EVib}).

\begin{equation}
 E_{lattice} = 1.81957 \lambda E_{*} \label{eqn:ELattice}
\end{equation}

\begin{equation}
 E_{vib} = 1.85188 \lambda^{1/2}E_{lattice} \label{eqn:EVib}
\end{equation}

By using the three-dimensional WKB method the zero temperature pycnonuclear reaction rate is given by Equation~(\ref{eqn:SPVHRate}).
The upper line is assuming a static lattice approximation and the lower line give rates for a completely relaxed lattice approximation.
The SPVH rates calculated for this work use the static approximation.

\begin{equation}
  P_0 = \frac{\rho}{\mu_A}A^2Z^4S\left( \begin{matrix} 3.90 \\ 4.76 \end{matrix} \right)10^{46}\lambda^{4/7}exp\left[-\lambda^{-1/2}\left(
\begin{matrix}
2.638 \\
2.516
\end{matrix}
\right)\right]
\label{eqn:SPVHRate}
\end{equation}

\section{Schramm-Koonin Rates from 1991}
In 1991 Schramm and Koonin \cite{SCHRAMMKOONIN1991} expanded on the work of SPVH to include Bravais lattices that are face centered ($\rm{fcc}$), and to take into account lattice polarization.
The expression given here is for $\rm{C^{12}-C^{12}}$ reactions only.
\begin{equation}
  P_0 = \left( \begin{matrix} 1.06 & \rm{bcc} \\ 2.69 & \rm{fcc} \end{matrix} \right)10^{45}S\rho A^2 Z^4\lambda^{4/7} \rm{exp}\left[-\alpha_2-\alpha_1\lambda^{-1/2}\left(
\begin{matrix}
2.638 \\
2.516
\end{matrix}
\right)\right]
\end{equation}

In their work, Schramm and Koonin considered the static case, the relaxed case, and the Weigner-Seitz approximation.
In the static case they again assumed that all ions except the fusing ions are fixed in their positions.
One of the effects of this is that no polarization in the lattice occurs.
When lattice polarization does occur, they also considered the case where the nuclei being considered for fusion have ``relaxed'' to their lowest energy state.
Assuming that the nuclei exist in Wigner-Seitz cells yields the Wigner-Seitz (WS) approximation. 
This approximation results in the wave vector being set to zero, $\textbf{k}=\textbf{0}$.
This then replaces the lattice unit cells with a sphere of the same volume as those unit cells.
This allows the transformation of a three-dimensional partial differential equation to a set of ordinary differential equations.

\begin{table}[hbt]
  \centering
  \begin{minipage}{3.5in}
    \centering
    \caption{Parameters for SK Pycnonuclear Reactions Rates\label{tab:KSParameters}}
    \begin{tabular}{|c||c|c|c|c|}    \hline
      Approximation &	$\rm{bcc}\hspace{3pt} \alpha_1$  &  $\rm{bcc}\hspace{3pt} \alpha_2 $ & $\rm{fcc}\hspace{3pt} \alpha_1$ & $\rm{fcc}\hspace{3pt} \alpha_2$ \\ \hline \hline
      Static & 2.639 & -6.305  & 2.401 & -6.315 \\ \hline
      WS & 2.516 & -6.793  & 2.265 & -6.911 \\ \hline
      Relaxed & 2.517 & -6.754  & 2.260 & -6.923  \\ \hline
    \end{tabular}
  \end{minipage}
\end{table}

\begin{figure}[!htb]
  \centering
  \begin{minipage}{4.5in}
    \includegraphics[width=\linewidth]{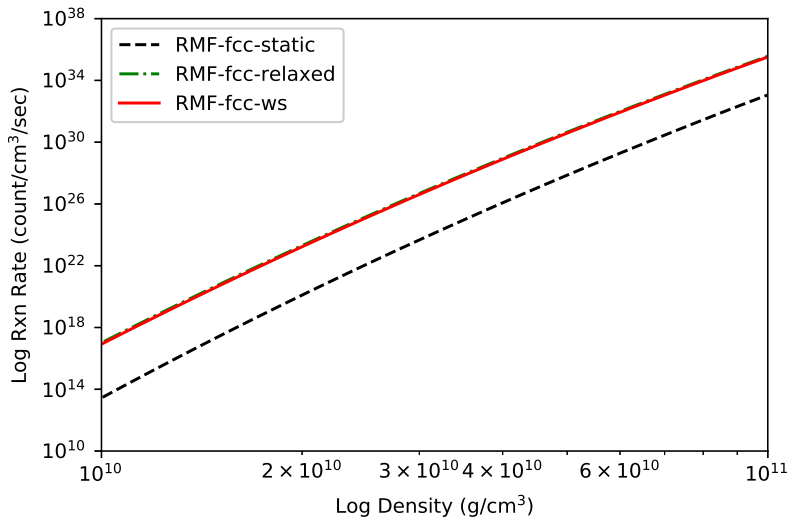}
    \caption{SK Rates for different lattice polarization approximations assuming RMF nucleon-nucleon interaction and fcc lattice structure.\label{fig:SKRateApproxComparison}}
  \end{minipage}
\end{figure}

For the SK reaction rates we can see from Figure~\ref{fig:SKRateApproxComparison} that the WS and relaxed approximations are almost identical.
This is due to the offsetting effects of lattice polarization and a change in the effective mass of the nuclei.

\begin{figure}[!htb]
  \centering
  \begin{minipage}{4.5in}
    \includegraphics[width=\linewidth]{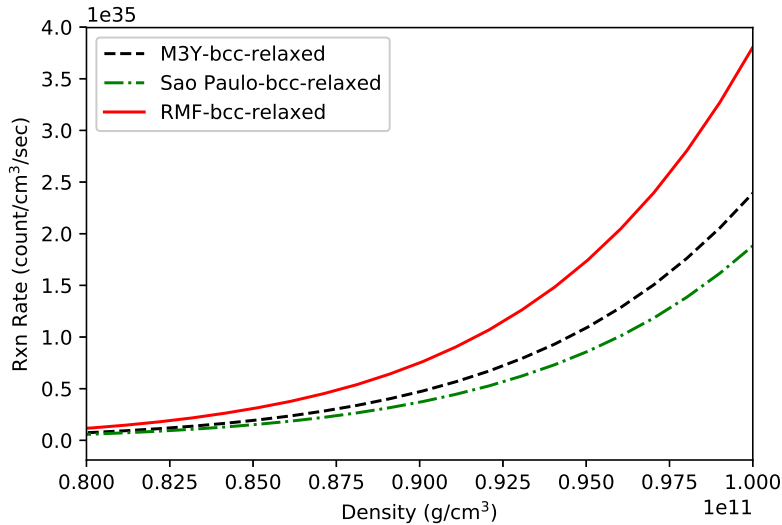}
    \caption{Schramm and Koonin rates for the three nucleon-nucleon interactions we are considering assuming a bcc lattice structure and relaxed lattice polarization approximation.\label{fig:SKRateNNComparison}}
  \end{minipage}
\end{figure}

\section{Analytical Rates from Gasques, et al. 2005}

In G05 the authors derived a set of analytical expressions for one component nuclear reactions for several physical regimes, including pycnonuclear burning and temperature enhanced pycnonuclear 
burning.
They applied this to study detonation of $\rm ^{12}C$-$\rm ^{12}C$ burning at densities $\rm \rho > 10^9\thinspace g/cm^3$.
The computer code to calculate these pycnonuclear reaction rates was not developed as part of this thesis, but was already available in the MESA stellar simulation software package.
Hence we will not delve to deeply into the details, but will make a few observations.

The reaction rates are formulated using the S\~{a}o Paulo nucleon-nucleon interaction.
A reason given in G05 for using this nucleon-nucleon interaction is that it contains no free parameters when nuclear density functionals are determined.
As can be seen in Figure~\ref{fig:nnpotcompare}, this potential does not have the hard-core feature that experimental evidence requires.
However, as can be seen in Figure~\ref{fig:SKRateNNComparison}, for the maximum densities we expect in our modeling of accreting white dwarf evolution, the pycnonuclear reaction rate calculated
assuming the S\~{a}o Paulo potential is within an order of magnitude of the rates calculated based on the M3Y and RMF nucleon-nucleon interactions.

In G05 temperature enhancement is automatically added in the appropriate density-temperature regime.
In Chapter~\ref{chapter:tempAndCrystal} and Appendix~\ref{appendix:compDetails} we will discuss some of the issues presented when trying to add temperature enhancement to the SPVH and SK reaction rate 
calculations.
Temperature enhancement of pycnonuclear reactions is still poorly understood, and is therefore difficult to calculate.

A relatively large part of our work for this thesis involved calculating the S-factor.
In G05 second order non-linear (NL2) effective interaction results were used to construct an analytical solution that is fit to experimental data.
The authors in G05 report that uncertainty in the NL2 effective interaction method is uncertain by a factor of $\sim 3.5$.
Theoretically our method here can be used to calculate the S-factor more exactly, although our calculation is much slower.

\begin{figure}[!htb]
  \centering
  \begin{minipage}{4.5in}
    \includegraphics[width=\linewidth]{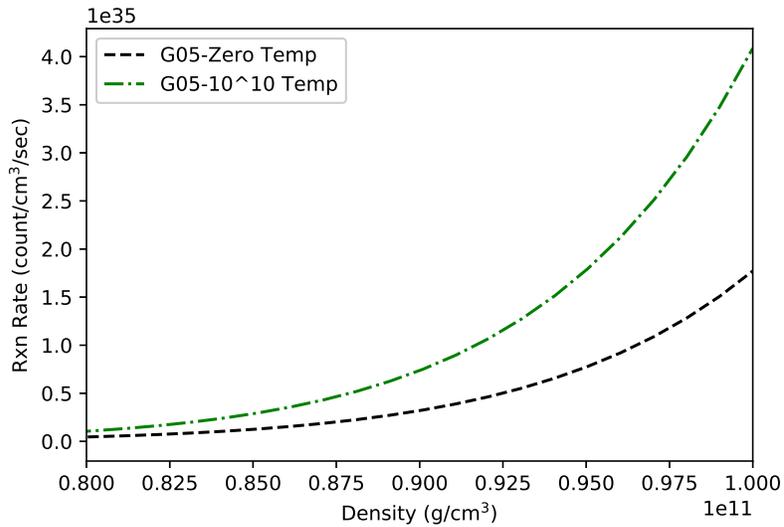}
    \caption{G05 reaction rates show a temperature dependence that is a result of changes to the S-factor with respect to temperature.\label{fig:graph}}
  \end{minipage}
\end{figure}

\section{Triple $\alpha$ Pycnonuclear Reaction Rates}

\begin{figure}[!htb]
  \centering
  \begin{minipage}{4.5in}
    \includegraphics[width=\linewidth]{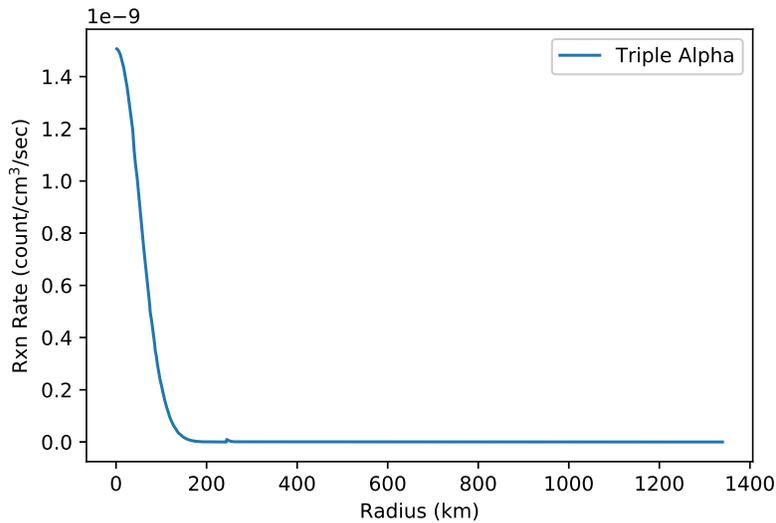}
    \caption{G05 triple-$\alpha$ particle reaction rates as a function of distance from the center of the star.\label{fig:g05finalTripleAlpha}}
  \end{minipage}
\end{figure}

\begin{figure}[!htb]
  \centering
  \begin{minipage}{4.5in}
    \includegraphics[width=\linewidth]{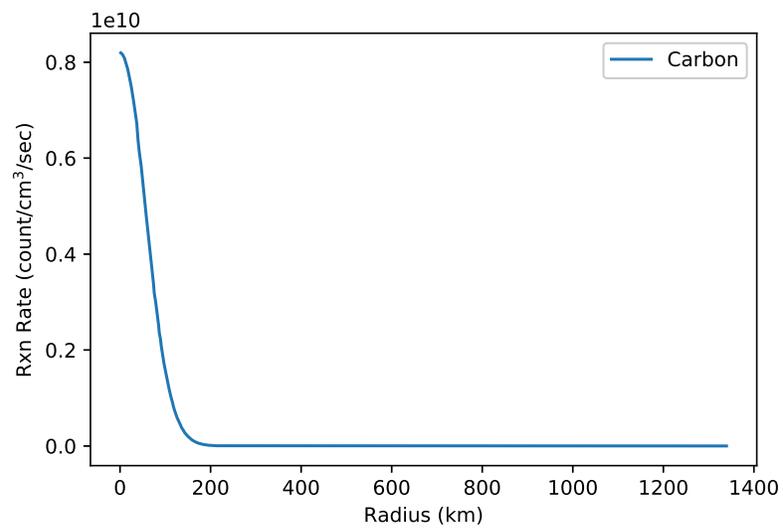}
    \caption{G05 carbon-carbon reaction rates as a function of distance from the center of the star.\label{fig:g05finalC12}}
  \end{minipage}
\end{figure}

Salpter and Van Horn derived expressions for the triple-$\alpha$ pycnonuclear reaction rate.
In 1987 Fushiki and Lamb \cite{FUSHIKILAMB1987} used S-Matrix calculations to derive rates for triple-$\alpha$ reactions in all temperature and density regimes.
MESA includes the ability to use these pycnonuclear reactions using the Fushiki and Lamb calculations, and we included them in our modeling.
However, as can be seen in Figures~\ref{fig:g05finalTripleAlpha} and ~\ref{fig:g05finalC12}, the triple-$\alpha$ rate is insignificant compared to the carbon reaction rate.

\chapter{FINITE TEMPERATURE RATES}
\label{chapter:tempAndCrystal}

The SPVH and SK pycnuclear reaction rates we have calculated thus far are for zero temperature.
In the case of white dwarfs zero temperature pycnonuclear reaction rates are often applicable.
However, as a detonation event quickly approaches temperatures can reach a point where temperature induced enhancement of reaction rates can occur.
This chapter explores how temperature may affect pycnonuclear reaction rates for the SPVH and SK calculations.

\section{Crystallization Condition}

Thus far in our discussions and formulations, we have assumed a lattice structure of body centered or face centered cubic unit cells.
It is reasonable to question if this a physically sound understanding of the micro-structure of the nuclei in the systems we are exploring.
Our reaction rates are assumed to be primarily based on density, but has the material in the white dwarf really formed into an ionic lattice?

One way to examine this issue is to look at the melting point of the material in the white dwarf.
This material will be crystallized below the melting temperature \cite{BHWD1983}.

\begin{equation}
  T_m \simeq \frac{Z^2 e^2}{\Gamma k}\left( \frac{4\pi}{3} \frac{\rho}{2Zm_u}\right)^{\frac{1}{3}}
\end{equation}

Here $\Gamma$ is defined as the ratio of the electrostatic energy to the thermal energy.  For $\Gamma \ll$ 1 a gas follows a nearly Maxwell-Boltzmann distribution.  For $\Gamma \gg$ 1 the plasma fluid crystallizes.
For crystal lattices in white dwarfs the estimated values for $\Gamma$ vary between 75 and 175 or so.

Assuming a $\mu_e \approx 2$, we can derive the simplified expression for melting temperature in Equation~(\ref{eqn:melting}).

\begin{equation}
  T_m \simeq 2\times 10^3\left(\frac{75}{\Gamma}\right)\rho^{1/3}Z^{5/3} K \label{eqn:melting}
\end{equation}

For material above this temperature, there may still be some lattice structure remaining distributed about any given volume element.
As the temperature increases the lattice will completely dissolve
The temperature at which this happens, $T_g$, based on Lindemann's Empirical Rule \cite{BHWD1983} is 16 times the melting point.

\begin{equation}
  T_g \simeq 3.2\times 10^4\left(\frac{75}{\Gamma}\right)\rho^{1/3}Z^{5/3} K
\end{equation}

\begin{figure}[!htb]
  \centering
  \begin{minipage}{4.5in}
    \includegraphics[width=\linewidth]{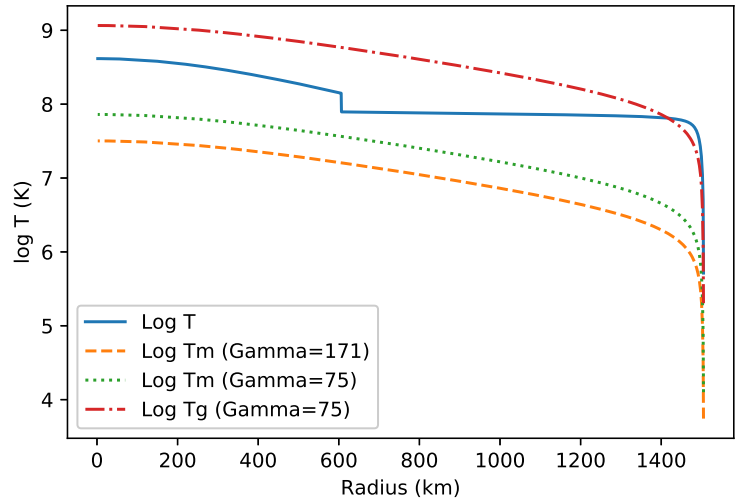}
    \caption{The melting points of the lattice crystal, the actual modeled temperature, and $T_g$ for the model with no pycnonuclear reactions.\label{fig:TandMeltTnopycno}}
  \end{minipage}
\end{figure}

We can see from Figure~\ref{fig:TandMeltTnopycno} that the modeled temperature of all zones of the accreting white dwarf simulation have $T\thinspace >\thinspace T_m$.
However the temperature is below $T_g$ indicating some lattice structure may still be present.
This indicates that in our models the lattice structure will be at least partially dissolved, and the rates we use may only be partially applicable in this model.
We leave further exploration of this matter for subsequent work.

\section{Temperature Enhancement}

In order to examine the effects of lattice vibrations SPVH introduced a dimensionless excitation parameter.

\begin{equation}
\beta=\frac{\left[\left(Z1+Z_2\right)^{5/3}-Z_1^{5/3}-Z_2^{5/3} \right]}{\left[Z_1^2Z_2^2A_1A_2/\left(A_1+A_2\right) \right]^{1/3}}\left(\frac{4.2579\times 10^7}{T}\right)\left(\frac{\rho/\mu_e}{1.6203\times 10^{10}} \right)^{1/3}\label{eqn:excitationparam}
\end{equation}

SPVH derives a rather complicated expression for the finite temperature enhancement to the pycnonuclear reaction rates.

\begin{equation}
  \frac{P}{P_0}=1+\left( \begin{matrix} 0.0430 \\ 0.0485 \end{matrix} \right)\lambda^{-\frac{1}{2}}
	\left[\left( \begin{matrix} 1.2624 \\ 2.9314 \end{matrix} \right)e^{-8.7833\beta^{3/2}}\right]^{-\frac{1}{2}}e^{f(\lambda,\beta)}\label{eqn:tempadjust1}
\end{equation}

\begin{equation}
  f(\lambda,\beta) = -7.272\beta^{\frac{3}{2}}+\lambda^{-\frac{1}{2}}\left( \begin{matrix} 1.2231 \\ 1.4331 \end{matrix} \right)
	e^{-8.7833\beta^{3/2}}\left[1-\left( \begin{matrix} 0.6310 \\ 1.4654 \end{matrix} \right)e^{-8.7833\beta^{3/2}}\right]\label{eqn:tempadjust2}
\end{equation}

\begin{figure}[!htb]
  \centering
  \begin{minipage}{4.5in}
    \includegraphics[width=\linewidth]{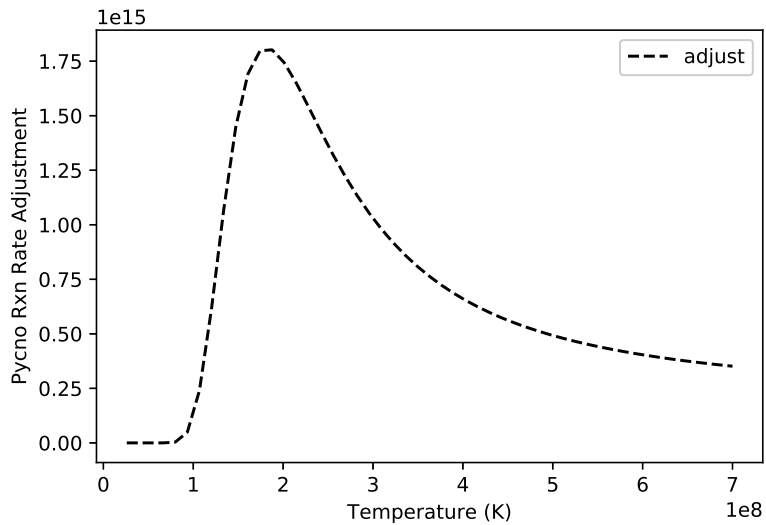}
    \caption{Enhancement of pycnonuclear reaction rate with respect to temperature at density $\rho=10^9$.\label{fig:RateTempAdjust9}}
  \end{minipage}
\end{figure}

Equation (\ref{eqn:tempadjust1}) gives a multiplier to the zero temperature reaction rate for a given temperature and density.
In Figure~\ref{fig:RateTempAdjust9} we show this temperature enhancement for white dwarf material at a density of $\rm \rho=10^9 \thinspace g/cm^3$.
It can be seen that the enhancement rises very quickly to a maximum and then stays fairly high up to temperatures of $ 7\times 10^8 \thinspace \rm{K}$ and beyond.
For the most part we will not include this temperature enhancement in our modeling.
However we will attempt to model an accreting white dwarf, with bcc lattice structure, M3Y nucleon-nucleon potential and the relaxed lattice approximation.
We expect we may have numerical difficulties due to the extremely rapid rise in temperature enhanced pycnonuclear reaction rates. 

\chapter{STELLAR ACCRETION ON TO A WHITE DWARF}

In this chapter we briefly describe some physical aspects of accretion onto a white dwarf.
For our simulations we ignore the effect of magnetic fields, thick versus thin disks, spin-up of white dwarf, and many other compilications.
Here we only give a brief description of the gravitation potential of close binary stars, the Chandrasekhar limit, and instability criteria.

\section{Basic Accretion Astrophysics}

Two stars that are in close orbit around a common center have the potential to transfer matter between them \cite{AccretionPower}.
In order to analyze this situation, it is common to assume that the stars are in circular orbits and to consider the reference frame that rotates with the major axis of the binary system.
Edouard Roche initially developed the theory for this motion.  
A Roche potential is defined to account for the centrifugal and Coriolis forces of the system.

\begin{equation}
\Phi_R(\textbf{r})=-\frac{GM_1}{|\textbf{r}-\textbf{r}_1|}-\frac{GM_2}{|\textbf{r}-\textbf{r}_2|}-\frac{1}{2}\left(\mathbf{\Omega} \times \textbf{r}\right)^2
\end{equation}

$M_1$ and $M_2$ are the masses of the binary component stars.
$r_1$ and $r_2$ are the distances from the origin to the first and second components of the of the binary. 
$\mathbf{\Omega}$ is the angular momentum of the binary system as it rotates with respect to an inertial reference frame.

\begin{equation}
\mathbf{\Omega}=\left[\frac{GM}{a^3}\right]^{1/2}\textbf{k}
\end{equation}

Here $M=M_1+M_2$ and $\textbf{k}$ is the unit vector perpendicular to the plane of motion.
$a$ is the distance between the stars.

\begin{figure}[!htb]
  \centering
  \begin{minipage}{4.5in}
    \includegraphics[width=\linewidth]{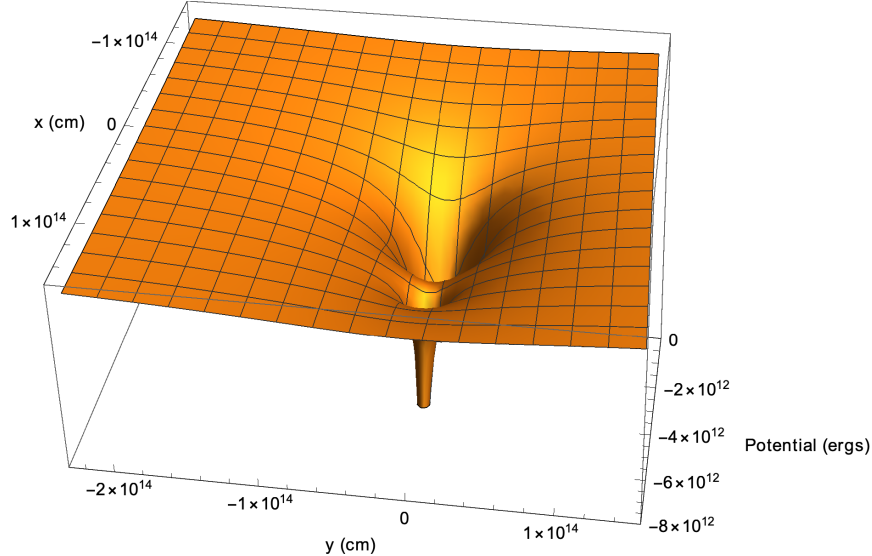}
    \caption{Roche potential with potential ``channel'' between the two stars located at the deep wells.\label{fig:RochePotential}}
  \end{minipage}
\end{figure}

As can be seen in Figure~\ref{fig:RochePotential}, there is a channel through which matter can flow.
During some portion of each of the stars' evolutionary path, expansion of the outer layers can occur which will result in mass transfer.

It is straightforward to show that astrophysical accretion processes can provide a great deal of power.
For a chunk of matter, $\Delta m$, falling on an object of mass M and radius $R_*$ the amount of energy released is $\Delta E_{acc}$.

\begin{equation}
\Delta E_{acc}=\frac{GM\Delta m}{R_*}
\end{equation}

For compact objects the amount of energy released per gram is between 20 times (for neutron stars) to about 0.25 (for white dwarfs) of the amount of energy released for nuclear reactions.
In addition, for both there is the possibility of bursting behavior as the amount of matter builds up and various types of thermonuclear runaway bursts start to occur.
By analyzing the luminosity of accretion and the expected black-body radiation from compact accretors, it can be shown that white dwarfs are expected to emit hard ultra-violet or soft x-ray radiation
and that neutron stars and black holes will emit hard x-rays and $\gamma$-rays \cite{AccretionPower}.

White dwarfs typically have magnetic fields on the order of $\rm 10^7 \thinspace G$.  
When material is transferred from a donor star to a magnetized white dwarf, the in-falling accretion disk, which is at least partially ionized, is disrupted with the material falling onto the magnetic 
poles.
This in-falling matter is organized into accretion columns.
These accretion columns can include shocks and are important for observational properties of the emitted x-rays of the binary system, but the effect on the white dwarf itself is, in essence, the same
as if the accretion were spherically symmetric.
For these reasons and since MESA is a one-dimensional stellar modeling application, we ignore this accretion of matter onto the magnetic poles, assuming that the accreted matter is uniformly distributed 
on the surface of the white dwarf.

\section{Chandrasekhar Limit}

By analyzing the total energy of a white dwarf, including internal energy and gravitational energy, as well as changes in those energies, one can use degenerate polytropic equations of state
to find a maximum mass, the Chandrasekhar limit. 
Correcting for general relativity yields the modern expression in Equation~(\ref{eqn:ChandrasekharLimit}).

\begin{equation}
M_{max}=1.457\left(\frac{\mu_e}{2}\right)^{-2}M_\odot \label{eqn:ChandrasekharLimit}
\end{equation}

In our modeling we assume a starting mass of one solar mass.
We accrete matter for approximately 400 million years at a rate of $10^{-9}\thinspace M_\odot$ and $\mu_e$ is very nearly 2 throughout our model.
If any of our final models have stars with a mass more than 1.457 $M_\odot$ we certainly have an inaccurate simulation.
      
\section{Onset of Instability}

In this work we used two different stopping criteria to complete our simulation
Both of these criteria were selected to indicate that the system is getting dynamic enough that changes to the observable properties of the white dwarf may occur.
The first of these was a very rough order of magnitude amount of energy released by nuclear fusion reactions, $\epsilon_{nuc} > 10^8 \thinspace \rm{ergs}$.
The second was based on a maximum temperature for any zone in the accreting white dwarf.
              
A detailed analysis provided in \cite{TempLimitRef} uses an analytical model that indicates that carbon ignition for a type Ia supernova should occur when central temperature is between $7.7-8.7 \times 10^8 \thinspace \rm{K}$.
The situation is actually a bit more complicated in that suspected instability occurs in regions $\rm 100-150 \thinspace km$ from the center of the white dwarf's center.
The reason this occurs is thought to be due to the convective nature of the material at the radial distances from the center.
The criterion derived for ignition is that Equation~(\ref{eqn:convPath}) diverges.

\begin{equation}
\int \left[ \left(\frac{dT}{dr}\right)_{exp}+\frac{\dot{S}_{nuc}}{c_Pv_{rms}} \right] dr \label{eqn:convPath}
\end{equation}

This integral is computed along a convective path.

Analyzing adiabatic heat flow leads to an expression for the derivative of temperature with respect to radius.

\begin{equation} 
\left(\frac{dT}{dr}\right)_{exp} \approx -0.037 T_c (\frac{\rho_9}{2})^{2/3}r_7
\end{equation}

In \cite{TempLimitRef} there is a discussion of various values for the heat capacity and possible $v_{rms}$ values.
In our study we will use a central temperature of $8.2\times 10^8 \thinspace \rm{K}$ for our cutoff.
However we will look at the convective regions close to the center and see if there are zones where the temperature exceeds $8.5 \times 10^8 \thinspace \rm{K}$. 

\chapter{MESA ENHANCEMENTS AND CONFIGURATION}
\label{chapter:mesaMods}

In order to perform the simulations desired we needed to modify and configure MESA.
Here we discuss the enhancements to MESA and then discuss the specific MESA configurations we used in our simulations.

\section{MESA Enhancements}

In order incorporate our pycnonuclear reaction rates, we modified the program unit \texttt{pycno.f90}.
This program unit is where the implementations of the pycnonuclear G05 $\rm ^{12}C$-$\rm ^{12}C$ and triple-$\alpha$ reaction rates are performed.

In order to use our rates, we do not calculate them directly during the course of the simulation.
Instead we embed a precomputed cubic spline in \texttt{pycno.f90}.

On initialization of \texttt{pycno.f90} a small configuration file is read to determine which specific pycnonuclear reaction rate is going to be used.
One entry of the file indicates which nucleon-nucleon interaction to use: RMF, M3Y or S\~{a}o Paulo.
The second entry indicates which combination of lattice structure and lattice polarization approximation is going to be used.
The final entry indicates if temperature enhancement will be performed.

Also incorporated into \texttt{pycno.f90} is a 5-point numerical differentiation routine.
This routine is required because the derivative of the reaction rate with respect to temperature and density needs to be calculated for our reaction rates.

For this work we used MESA version r11554.  
In that and earlier versions issues with how reaction rate screening was done in MESA was causing code maintenance issues for the MESA development team.
Therefore code for the G05 pycnonuclear reaction rates was removed from \texttt{pycno.f90} in versions subsequent to r11554.

\section{MESA Accreting White Dwarf Configuration}

The reaction rates we calculate are purely for pycnonuclear reaction rates.
In reality, in certain regions of the accreting white dwarf, temperatures and densities will be such that thermonuclear reactions will also occur.
MESA has the ability to use a variety of nuclear reaction networks.
For our models we use the \texttt{approx21} network.
This reaction network includes more than 21 particles species, including neutrons, protons and several nuclei, the most important which are indicated in Table~\ref{tab:approx21}.
These represent the most important nuclei in these types of systems in terms of energy production.
It also limits the number of reactions available so that the simulation is not overly computationally costly.

\begin{table}[hbt]
  \centering
  \begin{minipage}{5.5in}
    \centering
    \caption{Important Species in the \texttt{approx21} Reaction Network\label{tab:approx21}.}
    \begin{tabular}{|c|c|c|}    \hline
		Species & & \\ \hline
		n & p & H \\ \hline
		$\rm ^4He$ &	$\rm ^{12}C$ & N \\ \hline
		O & Ne & Mg \\ \hline
		S & Si & Ar \\ \hline
		Ca & Ti & $\rm ^{48}Cr$ \\ \hline
		$\rm ^{52}Cr$ & $\rm ^{54}Cr$ & Fe \\ \hline
		$\rm ^{56}Ni$ &  &  \\ \hline
	\end{tabular}
  \end{minipage}
\end{table}


MESA has the capability of defining the ionization level of the stellar material for the HELM equation of state.
In our models we configure MESA to use completely ionized material in the HELM equation of state by indicating that any material with a temperature above $\rm 10^{-20}$ K is completely ionized.

It is possible to instruct MESA whether or not to evolve the radial velocity of the zones being generated for each profile.
Since we are very interested in how accreting white dwarfs evolve over time when the energetics of the star change very quickly, we will certainly allow MESA to evolve the radial velocity of 
the simulated zones of the white dwarf models we generate. 

Another capability MESA has is the ability to determine if convection occurs in the zones it simulates for stellar evolution.
It can incorporate convection into the core hydrodynamic solver so that convective mixing and convective heat transfer are used when continuing the evolution.
Once again, since we are interested in modeling highly dynamic behavior of accreting white dwarfs we turn this capability on.

We will assume no rotation for this work, and hence our models will only include one-dimensional effects.
As stated previously, we assume a constant accretion rate of $\rm 10^{-9} \thinspace M_\odot/year$ and a composition of 25\% carbon and 75\% oxygen.

During the course of simulations, the time step size is adjusted in order to meet the resolution requirements that are specified.
For most of our models here we use a minimum time step size of $10^{-4}$ years.
An exception is in the model in which we attempt to include temperature enhancement.
In this model we use a minimum time step size of $10^{-20}$.
This model actually fails to complete for a normal nuclear energy production rate and maximum temperature value due to numerical stability issues.
These issues are related to the inability for certain zones to reach convergence quickly.
This causes MESA to choose smaller and smaller time steps until the time step is less that $10^{-2}$ seconds at which point the simulation ceases without reach the nuclear generation threshold.

Also as discussed previously, it is possible to tighten the range of the plasma coupling parameter, $\Gamma$, from the default of 150 to 175.
In this study we did not change that default.
Recollect that $\Gamma$ is the ration of electrostatic energy to thermal energy and is therefore unit-less

\chapter{RESULTS AND ANALYSIS}

Here we present the results of the simulations we ran.
Our results indicate that there are no observable properties of the white dwarf at detonation that would allow us to constrain our pycnonuclear
rate calculation micro-physics assumptions.
However, we can see some differences in observables between the case with pycnonuclear reactions and the case without.

\section{Final Model Observational Expectations}
 
The simulations we performed produced many megabytes of data, tracking a host of stellar properties in the different radial zones of the white dwarf.
Here we present the expected values of our final models for radius, effective surface temperature and age.
In this case the age indicates the amount of time the white dwarf has been accreting matter.

\begin{table}[hbt]
  \centering
  \begin{minipage}{5.5in}
    \centering
    \caption{Final Properties of Models\label{tab:finalProperties}.}
    \begin{tabular}{|c|c|c|c||c|c|c|c|c|c|}    \hline
      Calc.Type & NN-Pot. & Cell & Approx & $R$ (km) &  $T_{eff} (K)$ & Time (Myr) \\ \hline \hline
		No Pycno & - & - & - & 1505 & 96153 & 389.1 \\ \hline
		G05 & S\~{a}o Paulo & bcc & Static & 1340 & 106612 & 394.2 \\ \hline
		SPVH & M3Y & bcc & Static & 1339 & 106627 & 394.3 \\ \hline
		SK & M3Y & bcc & Relaxed & 1351 & 105836 & 393.9 \\ \hline
		SK & M3Y & bcc & Static & 1339 & 106717 & 394.3 \\ \hline
		SK & M3Y & bcc & WS & 1352 & 105910 & 393.9 \\ \hline
		SK & M3Y & fcc & Relaxed & 1347 & 106225 & 394.0 \\ \hline
		SK & M3Y & fcc & Static & 1337 & 106859 & 394.3 \\ \hline
		SK & M3Y & fcc & WS & 1347 & 106138 & 394.0 \\ \hline
		SPVH & Sao Paulo & bcc & Static & 1338 & 106759 & 394.3 \\ \hline
		SK & Sao Paulo & bcc & Relaxed & 1349 & 106016 & 394.0 \\ \hline
		SK & Sao Paulo & bcc & Static & 1337 & 106849 & 394.3 \\ \hline
		SK & Sao Paulo & bcc & WS & 1349 & 106086 & 393.9 \\ \hline
		SK & Sao Paulo & fcc & Relaxed & 1347 & 106223 & 394.0 \\ \hline
		SK & Sao Paulo & fcc & Static & 1338 & 106748 & 394.3 \\ \hline
		SK & Sao Paulo & fcc & WS & 1344 & 106386 & 394.1 \\ \hline
		SPVH & RMF & bcc & Static & 1339 & 106692 & 394.3 \\ \hline
		SK & RMF & bcc & Relaxed & 1354 & 105839 & 393.8 \\ \hline
		SK & RMF & bcc & Static & 1338 & 106926 & 394.2 \\ \hline
		SK & RMF & bcc & WS & 1354 & 105880 & 393.8 \\ \hline
		SK & RMF & fcc & Relaxed & 1351 & 105913 & 393.9 \\ \hline
		SK & RMF & fcc & Static & 1338 & 106749 & 394.3 \\ \hline
		SK & RMF & fcc & WS & 1350 & 105925 & 393.9 \\ \hline	\end{tabular}
  \end{minipage}
\end{table}

From Table~\ref{tab:finalProperties}, it can be seen that the only appreciable difference between the models is between the situation with no pycnonuclear reaction rates and the rest.
Between the models with pycnonuclear reaction rates there does not seem to be more than a quite small percentage of variation between the different reaction rate calculation characteristics.
In order to perform a more quantitative analysis, we compute the percentage differences from the mean for each case.
These percentage differences are presented in Appendix~\ref{appendix:errorsFromMean}.

Based on these results, it does not seem that we can observationally determine which pycnonuclear reaction rate calculations might be occurring in common astronomical observations.
None of the values we can observe vary by more than a total of 1\%, leading to the conclusion that these differences cannot be observed.
However, if we look at the differences in radius and effective surface temperature between the model with no pycnonuclear reactions and the models with pycnonuclear reactions, we see differences of 
approximately 10\%.
These variations should be detectable.

\section{Internal Structure Expectations}

In this section we wish to examine some of the differences that our models may exhibit in the internal structure.
We assume here that, since the bulk observational properties of the models with pycnonuclear reactions enabled are virtually identical, that their internal structure is nearly identical as well.
Hence we will compare the model without pycnonuclear reactions to only one of the models with pycnonuclear reactions.
The model we chose is one that uses the Schramm-Koonin (SK) rate calculations, assumes a body-centered cubic lattice structure (bcc) and uses the relaxed approximation for lattice polarization.

\begin{figure}[!htb]
  \centering
  \begin{minipage}{4.5in}
    \includegraphics[width=\linewidth]{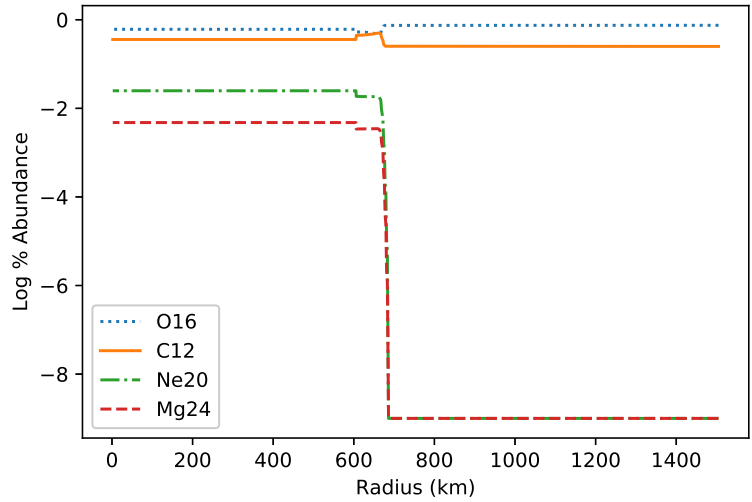}
    \caption{Final abundance of nuclei types for the model without pycnonuclear reactions.\label{fig:abundNopycno}}
  \end{minipage}
\end{figure}

\begin{figure}[!htb]
  \centering
  \begin{minipage}{4.5in}
    \includegraphics[width=\linewidth]{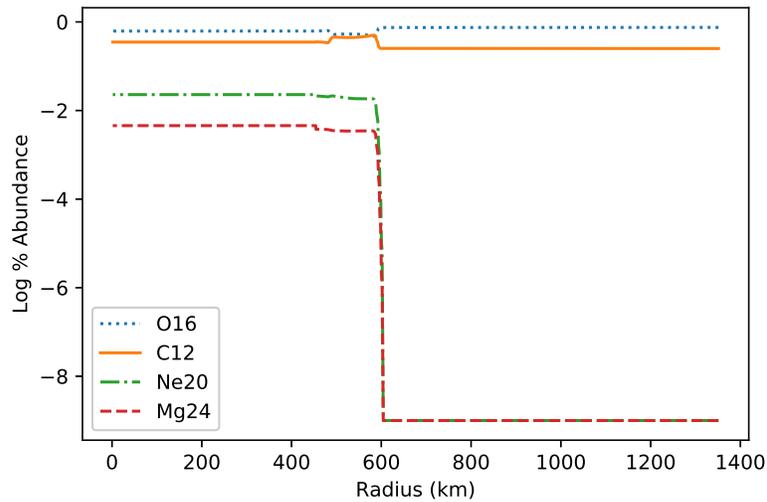}
    \caption{Final abundance of nuclei types for the model using M3Y nucleon-nucleon interaction, bcc lattice structure, and the relaxed approximation for polarized lattices.\label{fig:abundM3Ybccrelaxed}}
  \end{minipage}
\end{figure}

Comparing Figures~\ref{fig:abundNopycno} and~\ref{fig:abundM3Ybccrelaxed}, we can see that the only elements that are present in any appreciable quantities are $\rm{C}^{12}$, $\rm{O}^{16}$, $\rm{Ne}^{20}$, and $\rm{Mg}^{24}$ in both models.
Both models exhibit a step fall off in the abundances of $\rm{Ne}^{20}$ and $\rm{Mg}^{24}$.
For the pycnonuclear case, the fall off occurs at approximately 600 km, and for the case without pycnonuclear reactions it happens at 650 to 675 km.
The bottom abundances of the three low-abundance nuclei is approximately $\rm{10}^{-5}$ for the non-pycnonuclear reaction model and $\rm{10}^{-8}$ for the pycnonuclear reaction enabled model.

\begin{table}[hbt]
  \centering
  \begin{minipage}{5.5in}
    \centering
    \caption{Final Mass Fractions of Oxygen, Carbon, and Neon\label{wtab}.}
    \begin{tabular}{|c||c|c|}    \hline
      Nuclei & No Pycno & Pycno \\ \hline \hline
		O16 & 0.6406 & 0.6412 \\ \hline
		C12 & 0.3369 & 0.3383 \\ \hline
		Ne20 & 0.0174 & 0.0157 \\ \hline	
	\end{tabular}
  \end{minipage}
\end{table}

Even though the abundances of the various nuclei have distributions as a function of the distance from the center of the star the actual mass fractions of the only nuclei that have an appreciable 
abundance ($\rm{C}^{12}$, $\rm{O}^{16}$, and $\rm{Ne}^{20}$) do not vary by an amount that should be detectable.

As can be seen in Table~\ref{tab:TotalEnergyBySpecies}, the model with pycnonuclear reactions has a higher amount of power generated for nuclei 
whose reactions are less energetically favorable than carbon-carbon pycnonuclear reactions.
In particular the triple-$\alpha$ power decreases by two orders of magnitude, but the power is fairly low so it does not contribute much to the 
total energy budget of the system.
The nuclear power for this occurs even though we are including some pycnonuclear reactions for the triple alpha particle fusion energy as well.
The nuclear power liberated in the non-pycnonuclear reaction model is $\rm{1.41}\times 10^8 \thinspace L_\odot$ and for the model with pycnonuclear reactions we have a final nuclear 
power generated of $\rm{1.58}\times 10^8 \thinspace L_\odot$

\begin{table}[hbt]
  \centering
  \begin{minipage}{5.5in}
    \centering
    \caption{Log of Total Energy Released by Nuclear Species for Pycnonuclear Reaction Model and No Pycnonuclear Reaction Model\label{tab:TotalEnergyBySpecies}.}
    \begin{tabular}{|c||c|c|}    \hline
       & Log $\epsilon_{nuc}$ (ergs) &  Log $\epsilon_{nuc}$ (ergs) \\ 
      Nuclei & No Pycno & Pycno \\ \hline \hline
		pp & -19.61 & -19.90 \\ \hline
		CNO & -19.69 & -19.98 \\ \hline
		Triple Alpha & -5.57 & -7.43 \\ \hline
		N & -4.75 & -18.10 \\ \hline
		Ne & 11.12 & 11.38 \\ \hline
		Mg &  8.26 &  8.62 \\ \hline
		Si &  5.26 &  5.79 \\ \hline
		S &  3.78 &  4.28 \\ \hline
		Ar &  1.40 &  1.87 \\ \hline
		Ca & -0.72 & -0.25 \\ \hline
		Ti & -12.71 & -11.92 \\ \hline
		Cr & -26.67 & -25.35 \\ \hline
		Fe & -43.70 & -41.76 \\ \hline	
	\end{tabular}
  \end{minipage}
\end{table}

\section{Temperature Discontinuity}

As can be seen in Figures~\ref{fig:g05tempfinal} and~\ref{fig:nopcynotempfinal}, there is a discontinuity in the temperature curve for the final models.
This discontinuity is at just over 400 km for the case with pycnonuclear reactions and at approximately 600 km for the case without.

\begin{figure}[ht]
  \centering
  \begin{minipage}{4.5in}
    \includegraphics[width=\linewidth]{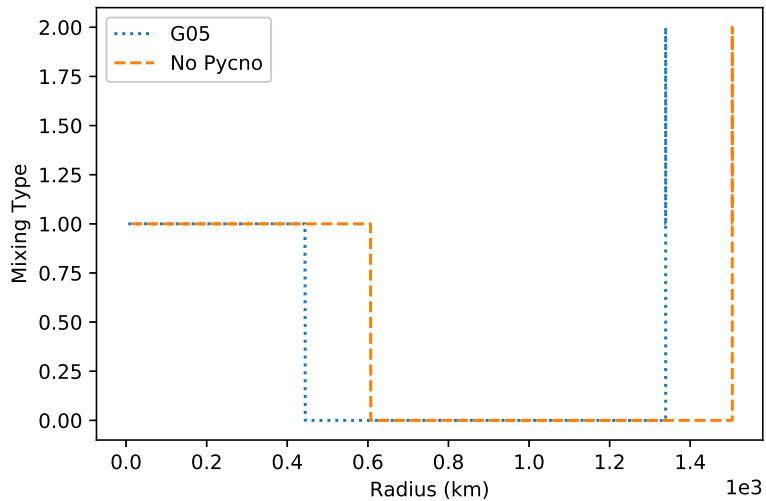}
    \caption{Mixing types for the final model of the G05 pycnonuclear reactions and the final model without pycnonuclear reactions.\label{fig:mixingTypes}}
  \end{minipage}
\end{figure}

In order to find the cause of this abrupt temperature drop, we reviewed several stellar properties, including pressure, density, opacity, convection type, and luminosity, all as a function of radius.
It is obvious from Figure \ref{fig:mixingTypes} that the temperature discontinuity is directly related to the convection boundary at $\sim 600$ km.
In MESA there are multiple types of mixing.
When the mixing type changes between two adjacent radial zones this indicates a convective boundary.
Table~\ref{tab:mixingTypes} lists the convective zones that occur in our models.

\begin{table}[hbt]
  \centering
  \begin{minipage}{5.5in}
    \centering
    \caption{MESA Mixing Types\label{tab:mixingTypes}.}
    \begin{tabular}{|c||c|}    \hline
      Mixing Type & Description \\ \hline \hline
		0 & No Mixing \\ \hline
		1 & Fully Convective \\ \hline
		2 & Overshoot \\ \hline
	\end{tabular}
  \end{minipage}
\end{table}

Comparing the values in this table to mixing types versus radius in Figure~\ref{fig:mixingTypes}, we see that we have convection inside the radial zone at around 600 km and then none until close to 
the white dwarf's surface.
It is hard to see it in the plot of mixing types, but when approaching the surface there are actually a few zones that have what is known as overshoot mixing, and then the rest of the zones that extend
to the white dwarf's surface are fully convective.
Convective overshooting here indicates that matter from the convective region near the surface is falling back toward the center of the white dwarf.
This region is only 7-8 meters thick.

\section{Evolution of Key Properties}

Thus far we have only been looking at the final models of our accreting white dwarf evolution simulations.
One of our goals was to also examine how the system evolves during the course of the simulation.
We still need to account for the larger radius and lower effective temperature at the surface which we noted in Table~\ref{tab:finalProperties}.
Examining how the accreting white dwarf evolves will shed some light into these differences.

All the figures presented in this section are done for the cases where no pycnonuclear reactions occur and using the G05 reaction rate calculations.
We looked at the difference between the G05 and the M3Y/bcc/relaxed SK model, and the differences were not appreciable.
Also, all the figures in this section use the model number of the start as the independent axis.
Recall that the time steps steps between model numbers is not consistent.
In these figures most of the highly dynamic changes in the properties of the star happen in the last million or so years.
By showing the properties by model number we can see the differences between the no pycnonuclear case and the case with pycnonuclear reactions.

\begin{figure}[ht]
  \centering
  \begin{minipage}{4.5in}
    \includegraphics[width=\linewidth]{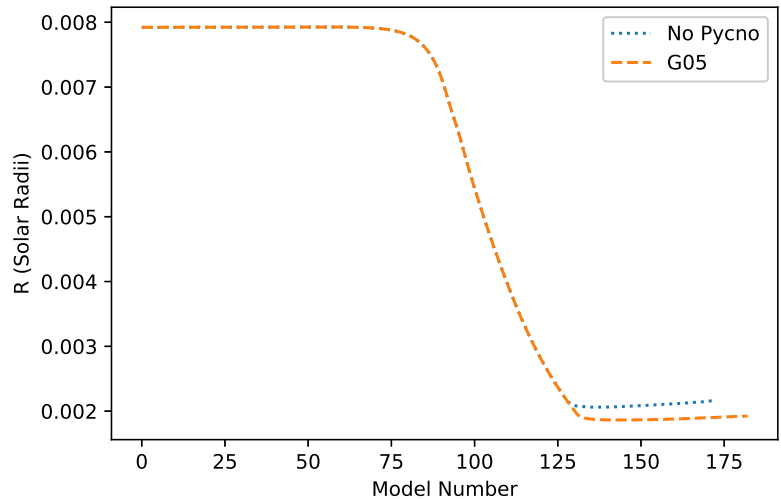}
    \caption{Evolution of the accreting white dwarf radius.\label{fig:evolveR}}
  \end{minipage}
\end{figure}

\begin{figure}[ht]
  \centering
  \begin{minipage}{4.5in}
    \includegraphics[width=\linewidth]{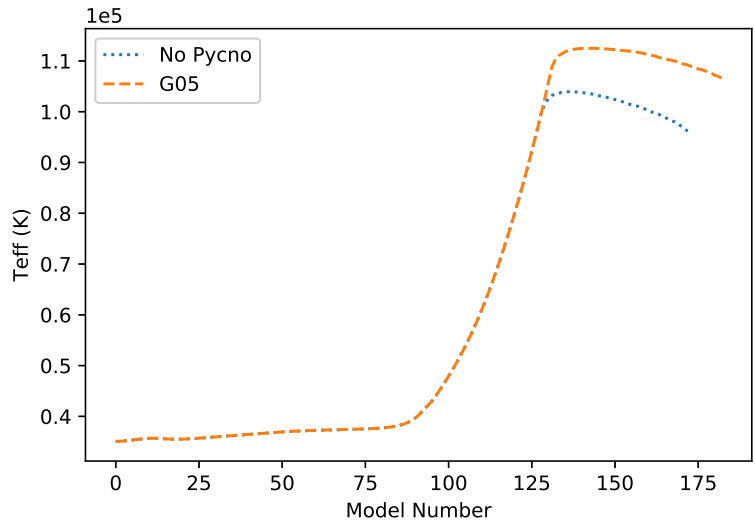}
    \caption{Evolution of the accreting white dwarf $T_{eff}$.\label{fig:evolveTeff}}
  \end{minipage}
\end{figure}

In Figure~\ref{fig:evolveR} we see that the radius for both cases goes down to a minimum and then starts to rise again.
Similarly Figure~\ref{fig:evolveTeff} shows $T_{eff}$ rising to a maximum coincident with the minimum radius, and then decrease somewhat before the model terminates.

\begin{figure}[ht]
  \centering
  \begin{minipage}{4.5in}
    \includegraphics[width=\linewidth]{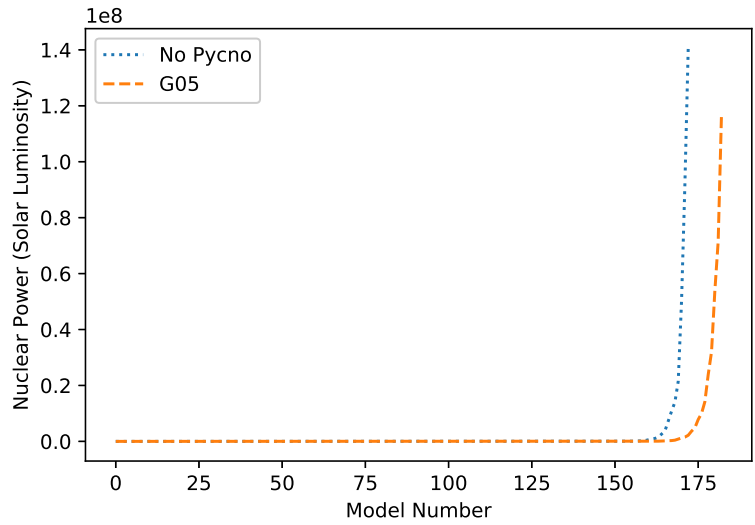}
    \caption{Evolution of the accreting white dwarf total nuclear power generated.\label{fig:evolveNucPower}}
  \end{minipage}
\end{figure}

By looking at the nuclear power generated in Figure~\ref{fig:evolveNucPower}, we can see that the total nuclear power generated in the no pycnonuclear reaction case rises at a lower model number than
for the G05 pycnonuclear reaction case.
It is uncertain why this occurs since we are removing a type of nuclear reaction, but we look at the white dwarf's carbon and neon mass as a function of model number to perhaps get some insight.

\begin{figure}[ht]
  \centering
  \begin{minipage}{4.5in}
    \includegraphics[width=\linewidth]{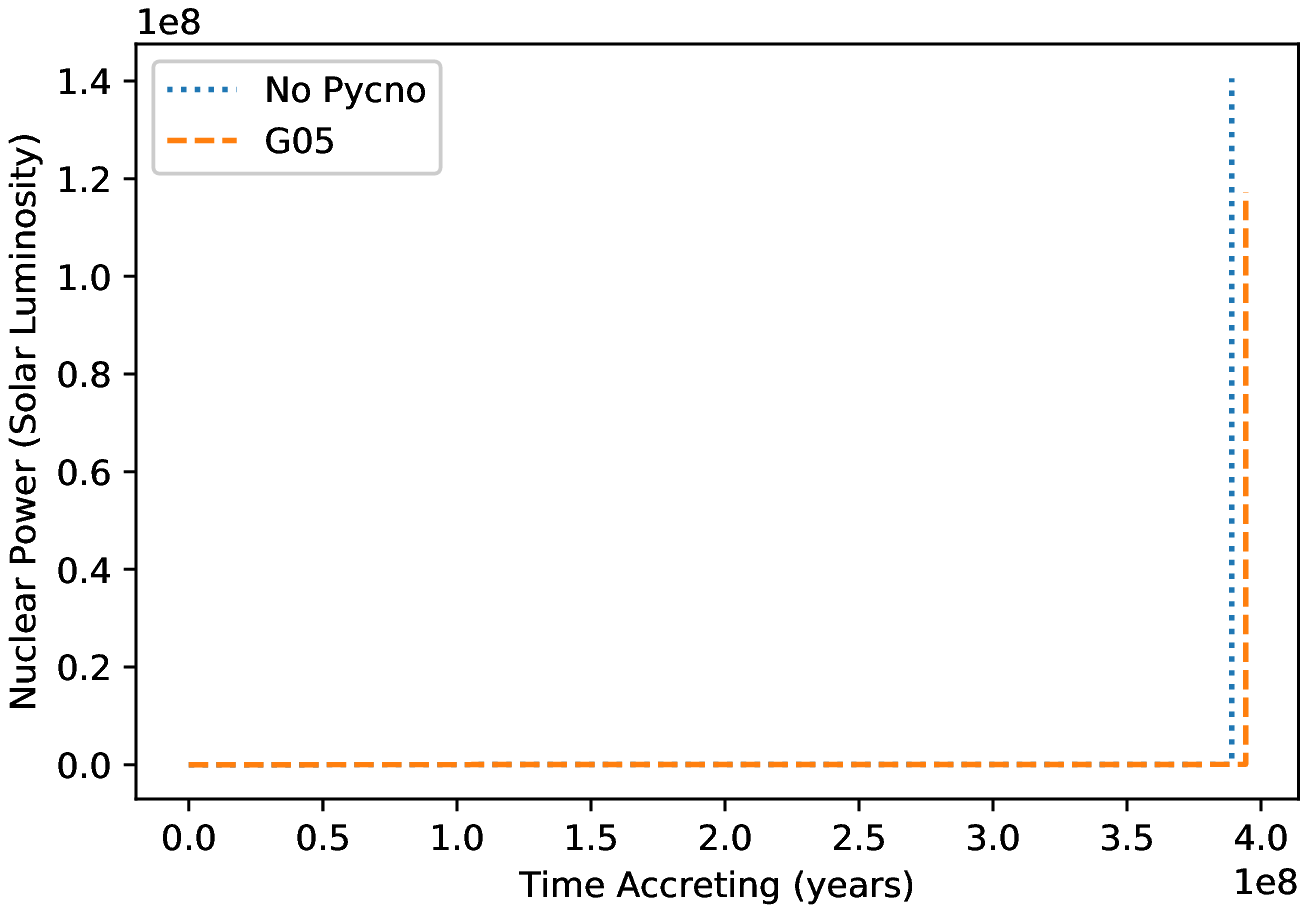}
    \caption{Evolution of the accreting white dwarf total nuclear power generated by accreting time.\label{fig:evolveNucPowerTime}}
  \end{minipage}
\end{figure}

\begin{figure}[ht]
  \centering
  \begin{minipage}{4.5in}
    \includegraphics[width=\linewidth]{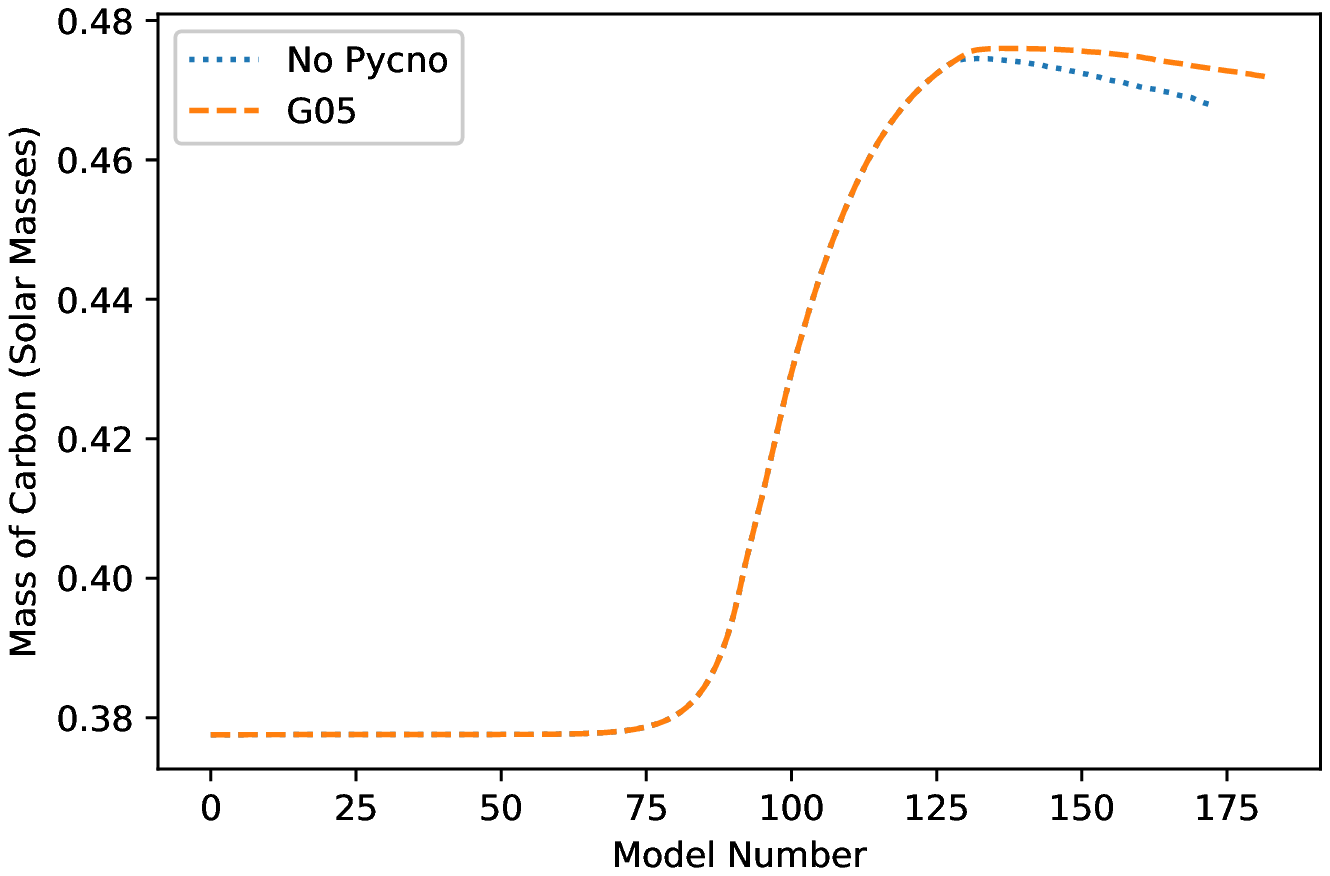}
    \caption{Evolution of the accreting white dwarf total carbon mass.\label{fig:evolveCMass}}
  \end{minipage}
\end{figure}

\begin{figure}[ht]
  \centering
  \begin{minipage}{4.5in}
    \includegraphics[width=\linewidth]{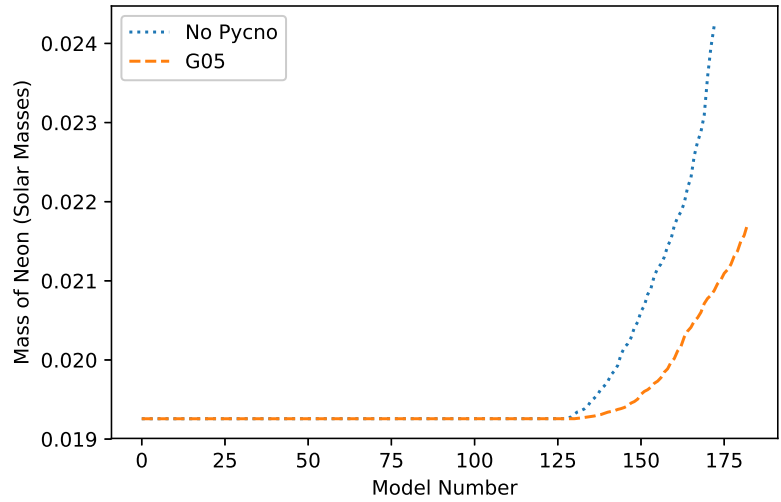}
    \caption{Evolution of the accreting white dwarf total neon mass.\label{fig:evolveNeonMass}}
  \end{minipage}
\end{figure}

In Figure~\ref{fig:evolveCMass} we see that the no pycnonuclear case is consuming carbon at a somewhat higher rate.
Looking at Figure~\ref{fig:evolveNeonMass} we see that neon is being created at a higher rate for the no pycnonuclear case than the case in which 
pycnonuclear reactions occur.
We do not show the evolution of the total oxygen mass here since it stays quite consistent throughout.
The total amount of mass for the other constituents is very low.

\begin{figure}[ht]
  \centering
  \begin{minipage}{4.5in}
    \includegraphics[width=\linewidth]{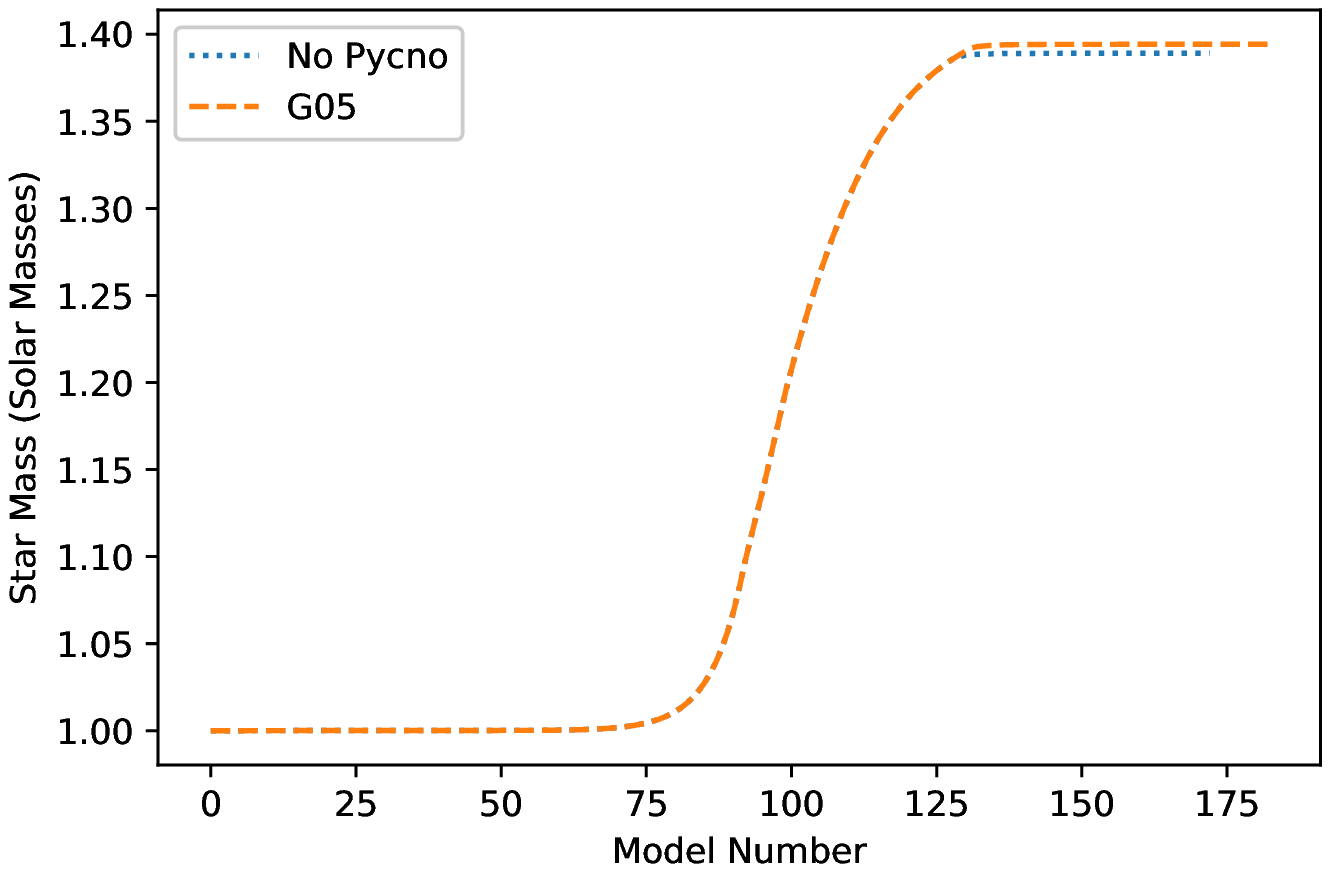}
    \caption{Evolution of the accreting white dwarf total mass.\label{fig:evolveMass}}
  \end{minipage}
\end{figure}

From these considerations we conclude that pycnonuclear reactions cause a lower rate of carbon burning than the case without pycnonuclear reactions.
Examining nuclear reaction networks can be very complicated and we make no attempts here to expound upon the particulars.
However, we present in Figure~\ref{fig:evolveMass} the evolution of the total white dwarf mass.
This figure shows that the case without pycnonuclear reactions has a slightly lower mass.
Since both models assume mass accretion and there is not appreciable stellar wind modeled, we can presume that this mass difference is due to the conversion of matter to energy by nuclear reactions.
White dwarf radii are inversely proportional to mass, hence the larger radius for the no pycnonuclear reaction case.
The lower temperature for the no pycnonuclear case is related to the lower density and temperature of the outer regions of the white dwarf undergoing accretion in that case.

\begin{figure}[ht]
  \centering
  \begin{minipage}{4.5in}
    \includegraphics[width=\linewidth]{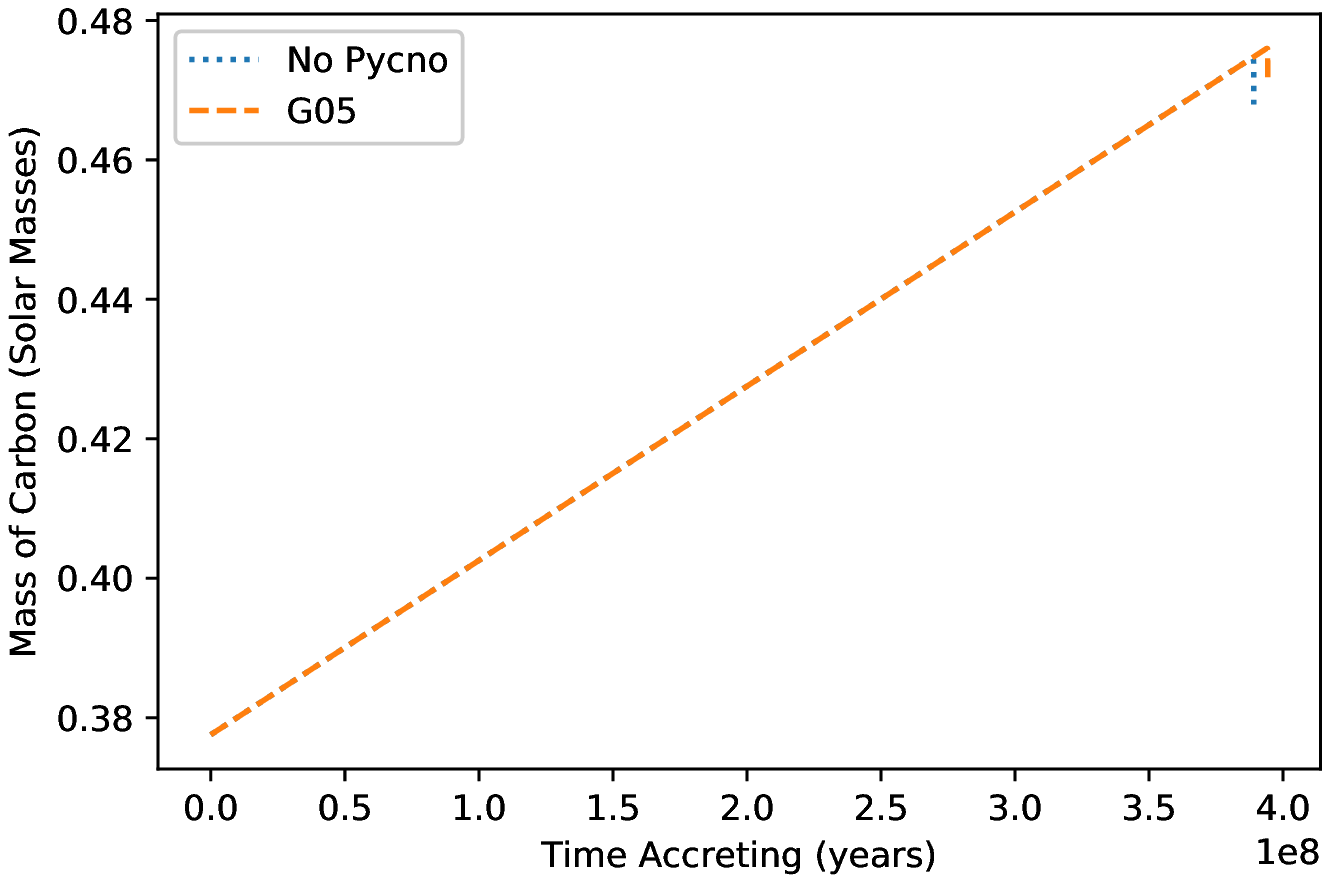}
    \caption{Evolution of the accreting white dwarf total carbon mass for time accreting.\label{fig:evolveCMassTime}}
  \end{minipage}
\end{figure}

\begin{figure}[ht]
  \centering
  \begin{minipage}{4.5in}
    \includegraphics[width=\linewidth]{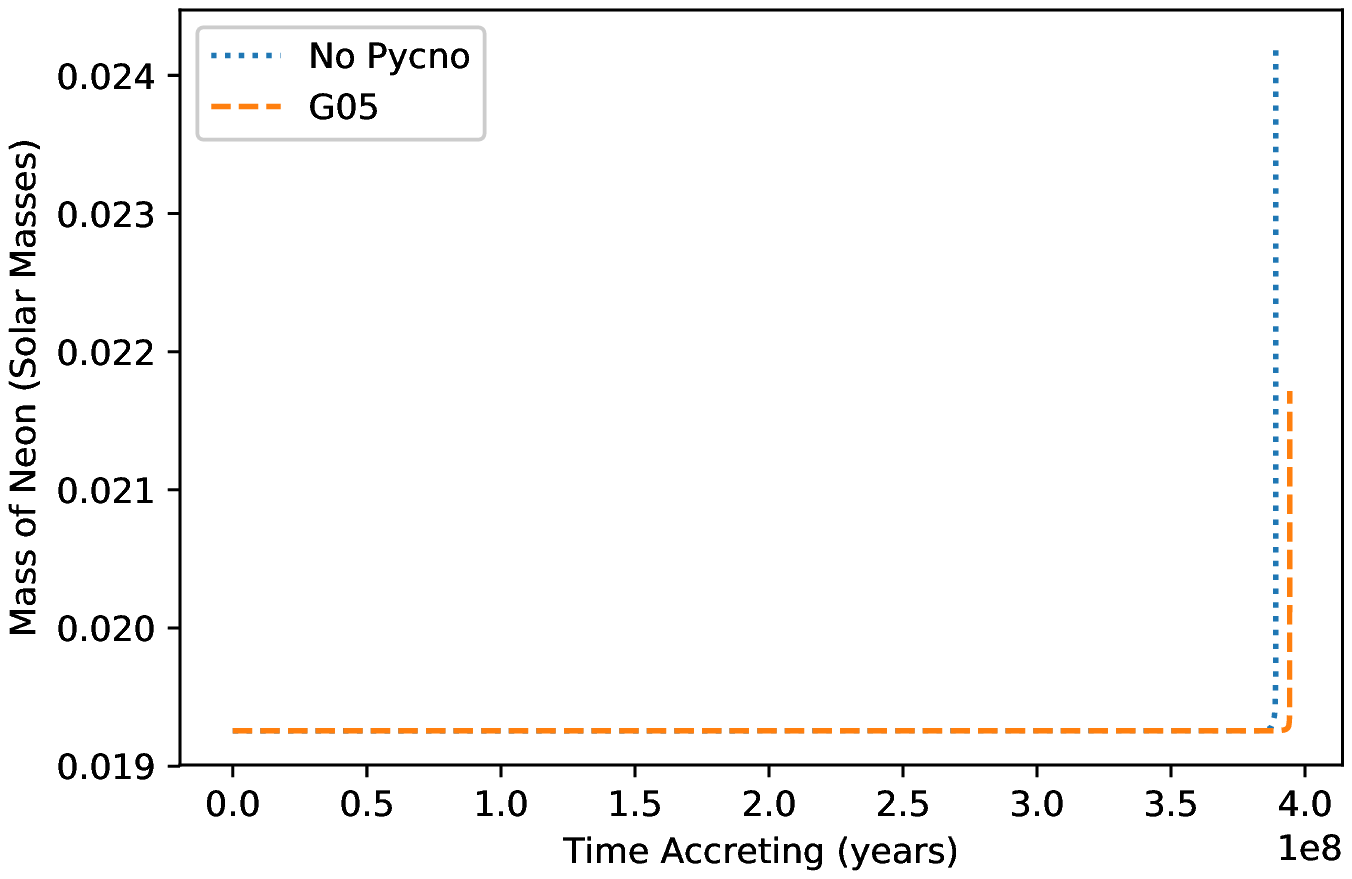}
    \caption{Evolution of the accreting white dwarf total neon mass for time accreting.\label{fig:evolveNeonMassTime}}
  \end{minipage}
\end{figure}

\begin{figure}[ht]
  \centering
  \begin{minipage}{4.5in}
    \includegraphics[width=\linewidth]{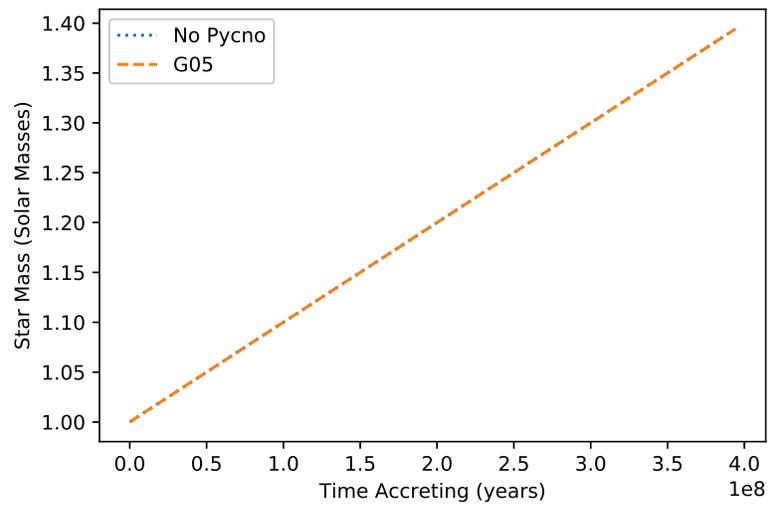}
    \caption{Evolution of the accreting white dwarf total mass for time accreting.\label{fig:evolveMassTime}}
  \end{minipage}
\end{figure}

Figures~\ref{fig:evolveCMassTime},~\ref{fig:evolveNeonMassTime},~\ref{fig:evolveMassTime}, and~\ref{fig:evolveNucPowerTime} show the Carbon Mass, Neon Mass, Total Mass, and Total Nuclear Power all as
functions of the accretion time.
As can be seen the disruptive event happens in a very short timescale highlighting a very quick detonation related to reaching the Chandrasekhar limit.

\chapter{CONCLUSION}

Here we give a short summary of our work and some ideas to expand on what we have done.

\section{Summary}

The primary conclusion we make from this numerical study is that the exact nature of the micro-physics of pycnonuclear rates does not yield observable variations in accreting white dwarf evolutionary 
simulations, at least not for the one-dimensional simulations we perform here.
In \cite{TempLimitRef} we find indications that, in order to get good results for type Ia supernovae ignition models for accreting white dwarfs close to the Chandrasekhar limit, full three-dimensional
methods must be employed.
Perhaps the same is true here.
In our models the central density only goes up to approximately $\rm 10^{10}\thinspace g/cm^3$.
These densities do not seem to significantly contribute the energy of the star, and the differences between the rates calculated assuming different micro-physics do not become noticeable until
the density approaches $\rm 10^{11} \thinspace g/cm^3$.
See Figure~\ref{fig:SKRateNNComparison} for example.

However, a more interesting finding is the higher final radius and lower effective temperature when we assume that no pycnonuclear reactions occur.
Therefore it seems conceivable that we could find a marker that would indicate whether these types of nuclear reactions do indeed take place.
Recalling that we hypothesize the possibility that non-crystallized white dwarf material will have significantly impacted pycnonuclear reaction rates this could be an important issue to explore.

\section{Future Work}

Here we list potential projects to further explore some of the subjects addressed in this thesis.
They are listed the in the order of our perceived importance.

\subsection{Three-Dimensional Accreting White Dwarf Evolution Simulation}

As discussed previously there are good reasons to believe that accreting white dwarf simulations need to be at least two-dimensional, and probably three-dimensional.
MESA is inherently a one-dimensional simulation tool.
It is possible to include rotation effect in MESA simulations, but it seems highly unlikely that this will yield adequate modeling of two-dimensional effects.
Of course these rotational capabilities will not address any sort of three-dimensional simulation effects.

\subsection{Apply Different Pycnonuclear Rate Calculations to Accreting Neutron Star Evolution}

The inner crusts of neutron stars have a density from around $4\times 10^{11}$ to $ \rm 5\times 10^{13} \thinspace g/cm^3$.
Hence the effects of pycnonuclear reactions should have considerably large effects on those systems.
Also, based on our calculations here, the variations in rates based on micro-physics assumptions, start to become greater at higher densities.

One of the complications from trying to model these types of systems is that the treatment for these stars should include general relativistic effects.
Also, as one approaches the core of the neutron star inverse beta-decay (so called neutron-drop) starts to alter the relative abundances of nuclei, and exotic nuclei start to form.
Some neutron stars have extremely high magnetic fields.
This can drastically alter screening of nuclear reactions, as well as cause nuclei to deform along the magnetic fields lines, perhaps excessively.

\subsection{Additional Nuclei Types}

In this work we have only included pycnonuclear reaction rates for one-component material consisting of $\rm ^{12}C$.
In order to do some of additional modeling we would like to, we need to include pycnonuclear reactions for other nuclei.
In addition it we would be well served if we could include multi-component materials as well to model $\rm ^{16}O$ reactions with $\rm ^{12}C$, for example.
The reaction rates calculated using the SPVH methods are easily extended to multi-component interactions.
However, the SK reaction rate methods need to be examined more closely in order to determine if the various integrations performed would modify any of the parameterizations we use.
Even for the SK reaction rates, the S-factor calculation is capable of being used for multi-component reactions.

\subsection{Pycnonuclear Reaction Rate Calculations}

The main component for calculating the folding potential, which is at the heart of generating the pycnonuclear reaction rates, already supports the interaction between different types of nuclei.
Indeed PycnoCalc can even calculate the folding potential between normal nuclei and strange quark matter (SQM) nuggets.  
An earlier version of the folding potential calculator was used to perform the calculations in \cite{GOLF2009}. 
The SPVH reaction rates calculated by us are easily extended to multi-component interactions.

Another potential variable in pycnonuclear reaction rates may involve the deformation of nuclei.
In highly magnetized white dwarfs and neutron stars it is possible for nuclei to become highly deformed, stretching out along the axis parallel to the magnetic field.
One possible change could be the folding potential between the nuclei: some parts of the nuclei would be closer, and other parts would be farther.
The exact effect on the folding potential would depend on the details of the deformation, as well as details of the nucleon-nucleon interaction.
It is difficult to evaluate the change to folding potential without actually performing the calculation.
It is worth noting that one of the reasons efforts have not been made to make the folding potential calculation of PycnoCalc faster is the desire to eventually perform the calculations using 
highly deformed nuclei.

Another effect that highly magnetized systems may have is related to potential changes in electron screening.

\subsection{Variable Accretion Rate and Composition}

Our models assume a constant accretion rate and accretion matter composition for nearly 400 million years.
This does not seems to be very realistic.
Using \texttt{MESABinary}, MESA can model accretion processes which include a variable accretion rate as well as changing accretion composition.
It may be possible to also model a variable accreted matter composition as well in this scenario.
A complication for doing this sort of modeling is that we would like to setup a configuration that produces accretion processes similar to the ones we have already done for this study, at least for a 
first effort.
Some inspired and educated guesswork maybe required, as well as a healthy dose of trial-and-error.

Another aspect we did not consider related to mass transfer is mass loss due to exceeding the Eddington luminosity limit and high stellar winds.
Further work should be done to examine this issue in more detail, and see how these effects may alter the models we generate here.

%
%
%
\bibliographystyle{siammod}
\bibliography{thbib}

\appendices
%
%

\chapter{PYCNOCALC}

We use the Fortran application PycnoCalc to perform all calculations for the pycnonuclear reaction rates in this thesis.
The complete source code is available in the Bitbucket repository at \url{https://bitbucket.org/sdsu_weber_research/pycnocalc/src/master}.
PycnoCalc was originally developed for the author's Physics Bachelor of Science degree thesis in 2008, where the routine to calculate the folding potential was originally created.
The folding potential routine of PycnoCalc was used for Master of Science theses of Golf and Ryan (n\'ee Currier) wherein pycnonuclear reaction rates were calculated for heavy ions with strange 
quark matter nuggets \cite{GOLFTHESIS,RYANTHESIS}.
This routine was also used to perform calculations reported in the Physics Review C article of Golf, Hellmers and Weber \cite{GOLF2009}.
Subsequent to the completion of those theses the calculations were incorporated into PycnoCalc.
We will now discuss some of the important Fortran modules and how they were used and modified to perform the calculations in this thesis.

\section{\texttt{nucleon\textunderscore interactions}}
This module contains the functions that calculate the M3Y, S\~{a}o Paulo, and RMF nucleon interactions.
The RMF interaction was added specifically for this thesis.  

This function for the RMF reaction supports the HS, Z, W and L1 parameter sets from \cite{SINGH2012,SINGH2012C} even though we only used the L1 parameter set in our calculations.

The $\delta$-function term in the M3Y nucleon-nucleon potential in Equation~(\ref{eqn:m3ypot}) is handled by an input parameter indicating the how close the volume elements of 
nuclear matter are allowed to get before simply adding the value $262$ to the calculated expression.

\section{\texttt{folding\textunderscore potential}}
This module contains the routine that calculates the folding potential.
This routine integrates over the volumes of both the target and the incoming nuclei, yielding a nested for loop of six levels.
It is very slow, however we have kept this structure in order to potentially compute the folding potential for deformed nuclei at a future date.
Thus far attempts to speed this routine up by using \texttt{OpenMP} parallelization have not been successful.
The only modification made for this thesis was to add the ability to use the RMF nucleon-nucleon interaction.

\section{\texttt{rate\textunderscore calc}}

The \texttt{rate\textunderscore calc} module contains functions to calculate the astrophysical S-factor, the SPVH, KS and G05 reaction rates.

\section{\texttt{csv\textunderscore kinds}, \texttt{csv\textunderscore module}, \texttt{csv\textunderscore parameters}, and \texttt{csv\textunderscore utilities}}

These modules are part of a comma-separated-value (CSV) formatted files open source package from Jacob Williams \url{https://github.com/jacobwilliams/fortran-csv-module}.
This package was incorporated into PycnoCalc in order to more efficiently write and read CSV files.
We use the Python package \texttt{pandas} to help with generating plots, as well as do other data analysis for PycnoCalc output, and \texttt{pandas} seamlessly uses CSV files.

\section{\texttt{mathdiff}, \texttt{mathintegration}, \texttt{mathinterpolation}, \texttt{mathlinearalg}, and \texttt{mathutils}}

These modules are primarily based on the algorithms and software described in \cite{NUMERICALRECIPES}.
The routines in these modules are not meant to be comprehensive or even particularly robust, but are implemented for quick, useful routines to perform the basic numerical analysis we need for PycnoCalc.
The module \texttt{mathintegration} implements the trapezoid quadrature method.
One version is based on function values preloaded into an array, and another version has a function name passed in and computes the numeric integral for that function.
These methods are used to calculate the 2pF nuclear density functional and the WKB approximation for the S-factor, respectively.
We implement the creation and the use of cubic splines in \texttt{mathinterpolation}.
This is used to calculate the pycnonuclear rates used in the \texttt{pycno.f90} MESA program unit.
In \texttt{mathlinearalg} we implement the solving of tridiagonal systems of equations.
This is needed in order to compute the cubic spline.
The 5-point numerical differentiation in module \texttt{mathdiff} is calculated as described in \cite{koonin1990computational}.

\section{\texttt{tests} and \texttt{system\textunderscore functions}}

The module \texttt{tests} contains all the unit tests that are done during development.
A configuration file name is passed into PycnoCalc, indicating which unit tests will be performed.
At this time we have no baselines to track against, but we can at least rerun tests with the correct parameters to verify basic functionality.

The module \texttt{system\textunderscore functions} contains all the code that actually produces research results that we are interested in retaining.
For example, the routines that generate the pycnonuclear reaction rates are contained in this module.
When the routines in this module execute, they typically output data into a format that we will use either in another piece of code, like \texttt{pycno.f90}, or to plot or
put that data into a \LaTeX\ table.
Similar to unit tests, there is a configuration file that is used for each function that generates research data.

\chapter{SOME COMPUTATIONAL DETAILS}
\label{appendix:compDetails}
\section{Use of Bitbucket for Source Code Management}

All of the artifacts generated while creating this thesis, except for the raw model data from the models and logs files, are managed in the Bitbucket repository.
Bitbucket provides free source code repositories with certain limitations.
The main limitations are only 5 users for private repositories, and the repository size cannot be more than 1 GB.

Source code management provides many benefits.
One benefit is the ability to have a central location for all elements required to reproduce the results of a particular project, this Master's thesis in this case.
Another benefit is the ability to merged incremental updates of the source code and other artifacts.
This allows for the creation of a consistent baseline at important milestones or task completions.
Very usefully source code repository users can revert to previously known good versions of the application

For larger software projects, using source code management allows multiple software developers to enhance and maintain a software application at the same time.
Typically developers create a branch of the software, and pull the branch onto a computer they use for development.
When completed, they can the push their local changes onto the remote branch in Bitbucket, and then they or another developer can merge those changes into the master branch.
Any conflicts between changes by different developers to the same artifacts can be resolved in this process.

Another major reason to use an online source code repository related to reproducibility is that the repository can act as a permanent record of how the research results were calculated.
It is possible that Bitbucket, which is operated by the company Atlassian, may someday become defunct, but it is also possible to download repositories into a generic \texttt{git} repository
on a file system, and then load it into another source code management system like GitHub or GitLab.

Two sets of artifacts that we are not able to keep in Bitbucket are the raw data for the models and the job output and error logs.
The reason for this is simply because they are too big, taking almost 800 megabytes when compressed.
This amount of data is not really appropriate for a source code management system.
This data is available upon request.

\section{SDSC Used to Generate Models}

All the models list in Tables~\ref{tab:microAssumptions} and~\ref{tab:additionalModels} were run as jobs on the San Diego Supercomputing Center (SDSC) supercomputer Comet.

\begin{table}[hbt]
  \centering
  \begin{minipage}{6.5in}
	\centering
  	\caption{Additional Models Generated\label{tab:additionalModels}}
    \begin{tabular}{|c|c|}    \hline
  		wd\textunderscore ignite\textunderscore nopycno & Accreting white dwarf with no pycnonuclear reactions  \\ \hline
		wd\textunderscore ignite\textunderscore m3y\textunderscore bcc\textunderscore relaxed\textunderscore adj & SK, M3Y, bcc, relaxed with temp. enhancement  \\ \hline
	\end{tabular}
  \end{minipage}
\end{table}

There were two types of jobs that were submitted for this thesis.
The first were the jobs to calculate the pycnonuclear reaction rates.
These jobs were computationally expensive primarily due to the non-optimized, non-parallelized way the double folding-potential is calculated.
There were three long running jobs that ran for nearly 24 hours each, one for each nucleon-nucleon potential.

The second type of job was one which ran the simulation of the accreting white dwarf.
Each type of model had a configuration stored in a folder bearing the model name.
Each job initialized and built the model's software for the run, cleared out data from previous runs, ran the simulation, moved the important model data to a separate folder, and deleted the data in
which we were not interested. 

For the temperature enhanced pycnonuclear reaction model, \texttt{wd\textunderscore ignite\textunderscore m3y\textunderscore bcc\textunderscore relaxed\textunderscore adj}, we were unable to achieve
convergence.
In our efforts to solve the problem we handled this job slightly differently.
For this model we set the number of \texttt{OpenMP} threads to 24 (the maximum allowed on Comet), and set the maximum duration for the job to 48 hours.
This is compared to the rest of the models which achieved convergence relatively quickly (less than an hour) with only 2 threads.

\section{Troubleshooting MESA Convergence Issues}

The models that were initially generated using pycnonuclear reaction rates associated with polarized lattice relaxed condition, or the Wigner-Seitz approximation, never actually reached the 
terminating condition for either energy output or a high enough temperature.
At first we felt this might be due to actual physical differences in the models.
However, it was soon discovered that the reason the models never reached our instability criteria was because the time step used by MESA became very small, less than one year.
Further analysis showed this was due to a large number of Newton-Raphson iterations required to reach stability for some of the shells of the models.
MESA has the characteristic that when the number of these iterations is larger than a certain value (14 by default), the time step will be reduced.
The models that did not have this problem did not have a time step less than around one million years.
In order to identify the zones where the problem was occurring, we used MESA debugging tools.
Using one of these tools we were able to generate a graphic which allowed us to narrow down the issues to a region.

\begin{figure}[!htb]
  \centering
  \begin{minipage}{4.5in}
    \includegraphics[width=\linewidth]{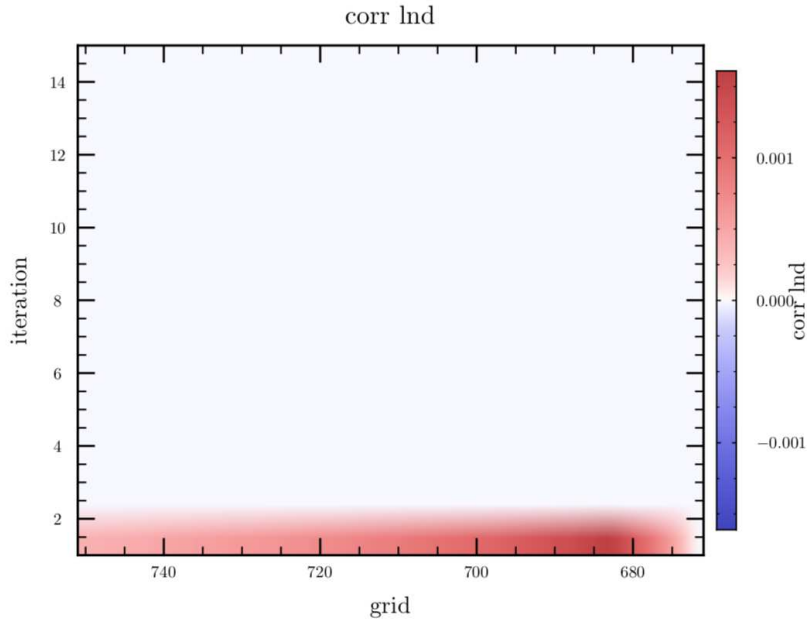}
    \caption{The red areas in this troubleshooting graphic indicate areas where the number of Newton-Raphson iterations is over the threshold to decrease the time step size.\label{fig:MESATroubleshooting}}
  \end{minipage}
\end{figure}

Further research indicated that on some noise in some of the pycnonuclear reaction rates, and very low values of those rates, was causing the spline we used to calculate negative reactions rates.
This problem was ultimately resolved by using a threshold below which reaction rates and their associated derivatives would be set to zero.
This threshold is set to $10^{-200}$.

\section{Lack of Convergence for Model Using Temperature Enhancement}

\begin{figure}[!htb]
  \centering
  \begin{minipage}{4.5in}
    \includegraphics[width=\linewidth]{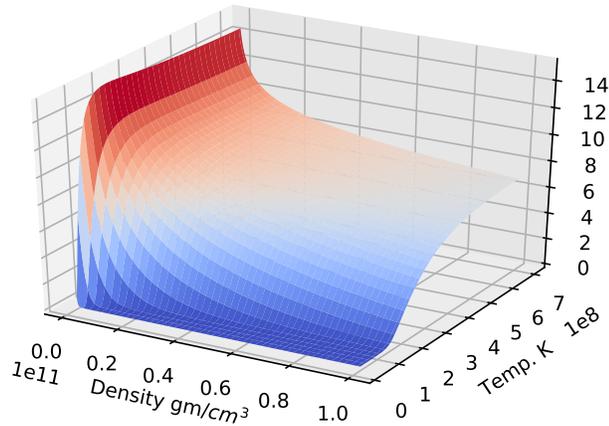}
    \caption{The Log of the temperature adjustment when using the static lattice approximation.\label{fig:tempAdjustSurfStatic}}
  \end{minipage}
\end{figure}

\begin{figure}[!htb]
  \centering
  \begin{minipage}{4.5in}
    \includegraphics[width=\linewidth]{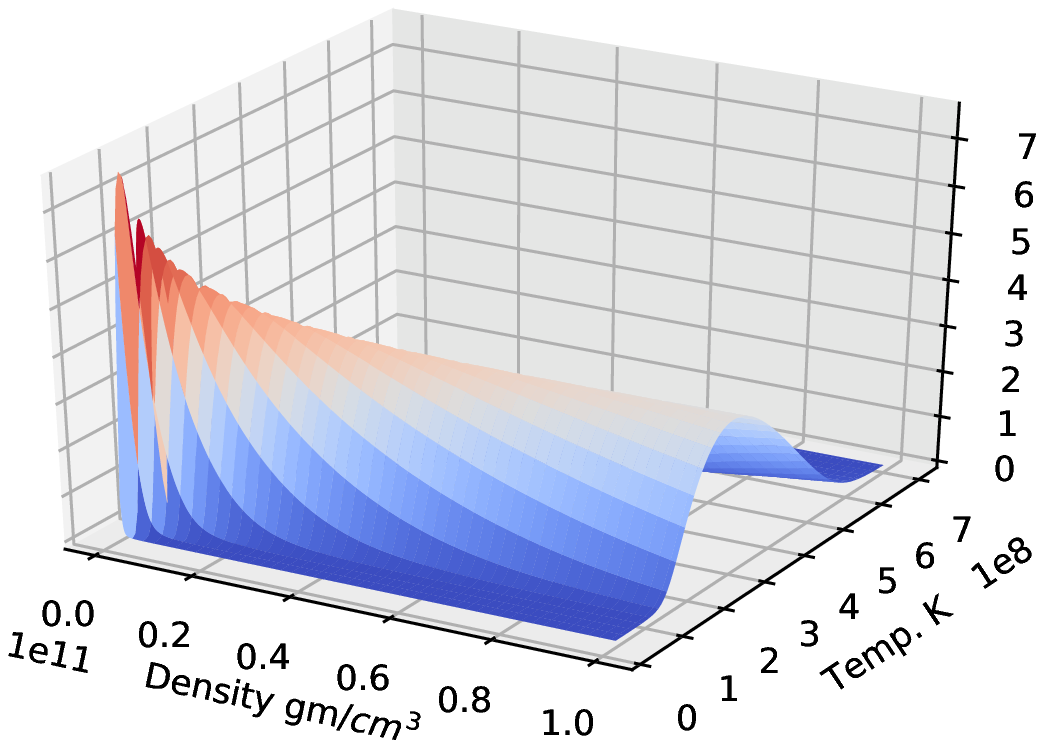}
    \caption{The Log of the temperature adjustment when using the relaxed lattice approximation.\label{fig:tempAdjustSurfRelaxed}}
  \end{minipage}
\end{figure}

One of the more challenging physical values to calculate is the adjustment of pycnonuclear reactions rates based on temperature.
It is tempting to simply perform the calculations in Equations~(\ref{eqn:excitationparam}),~(\ref{eqn:tempadjust1}), and~(\ref{eqn:tempadjust2}).
However, as shown in Figures~\ref{fig:tempAdjustSurfStatic} and~\ref{fig:tempAdjustSurfRelaxed}, there are features at the lower densities that are rather nonphysical
In order to deal with this situation, we have modified our implementation to only perform the temperature adjustment calculation for a relatively narrow range of densities and temperatures.
However, there was never enough numerical stability to allow the model to achieve the normal temperature or nuclear energy stopping conditions.

\chapter{PERCENT ERROR FROM MEAN OF OBSERVATIONAL PROPERTIES}
\label{appendix:errorsFromMean}
Tables~\ref{tab:radiiErrors} and~\ref{tab:teffErrors} show percentage deviations from the mean of the final radii and effective temperatures respectively for all of our different pycnonuclear rate calculations.
We have included negative percentages to indicate that the value is lower than the mean value.
In all cases the variation in expect values are less than 1\% and are not likely to be observationally detectable.

\begin{table}[hbt]
  \centering
  \begin{minipage}{5.5in}
    \centering
    \caption{Percent Error from Mean of Final Radii\label{tab:radiiErrors}.}
    \begin{tabular}{|c|c|c|c||c|c|}    \hline
      Calc.Type & NN-Pot. & Cell & Cell Approx. & R (km) & \% Error from Mean \\ \hline \hline
	No Pycno & - & - & - & 1505 & - \\ \hline
	G05 & S\~{a}o Paulo & bcc & Static & 1340 & -0.36 \\ \hline
	SPVH & M3Y & bcc & Static & 1339 & -0.42 \\ \hline
	SK & M3Y & bcc & Relaxed & 1351 & 0.51 \\ \hline
	SK & M3Y & bcc & Static & 1339 & -0.42 \\ \hline
	SK & M3Y & bcc & WS & 1352 & 0.55 \\ \hline
	SK & M3Y & fcc & Relaxed & 1347 & 0.17 \\ \hline
	SK & M3Y & fcc & Static & 1337 & -0.55 \\ \hline
	SK & M3Y & fcc & WS & 1347 & 0.20 \\ \hline
	SPVH & Sao Paulo & bcc & Static & 1338 & -0.49 \\ \hline
	SK & Sao Paulo & bcc & Relaxed & 1349 & 0.34 \\ \hline
	SK & Sao Paulo & bcc & Static & 1337 & -0.58 \\ \hline
	SK & Sao Paulo & bcc & WS & 1349 & 0.37 \\ \hline
	SK & Sao Paulo & fcc & Relaxed & 1347 & 0.17 \\ \hline
	SK & Sao Paulo & fcc & Static & 1338 & -0.50 \\ \hline
	SK & Sao Paulo & fcc & WS & 1344 & -0.02 \\ \hline
	SPVH & RMF & bcc & Static & 1339 & -0.42 \\ \hline
	SK & RMF & bcc & Relaxed & 1354 & 0.69 \\ \hline
	SK & RMF & bcc & Static & 1338 & -0.49 \\ \hline
	SK & RMF & bcc & WS & 1354 & 0.73 \\ \hline
	SK & RMF & fcc & Relaxed & 1351 & 0.52 \\ \hline
	SK & RMF & fcc & Static & 1338 & -0.44 \\ \hline
	SK & RMF & fcc & WS & 1350 & 0.43 \\ \hline
	\end{tabular}
  \end{minipage}
\end{table}

\begin{table}[hbt]
  \centering
  \begin{minipage}{5.5in}
    \centering
    \caption{Percent Error from Mean of Final Effective Temperatures\label{tab:teffErrors}.}
    \begin{tabular}{|c|c|c|c||c|c|}    \hline
      Calc.Type & NN-Pot. & Cell & Cell Approx. & Teff (K) & \% Error from Mean \\ \hline \hline
		No Pycno & - & - & - & 96153 & - \\ \hline
		G05 & S\~{a}o Paulo & bcc & Static & 106612 & 0.24 \\ \hline
		SPVH & M3Y & bcc & Static & 106627 & 0.25 \\ \hline
		SK & M3Y & bcc & Relaxed & 105836 & -0.49 \\ \hline
		SK & M3Y & bcc & Static & 106717 & 0.34 \\ \hline
		SK & M3Y & bcc & WS & 105910 & -0.42 \\ \hline
		SK & M3Y & fcc & Relaxed & 106225 & -0.13 \\ \hline
		SK & M3Y & fcc & Static & 106859 & 0.47 \\ \hline
		SK & M3Y & fcc & WS & 106138 & -0.21 \\ \hline
		SPVH & Sao Paulo & bcc & Static & 106759 & 0.38 \\ \hline
		SK & Sao Paulo & bcc & Relaxed & 106016 & -0.32 \\ \hline
		SK & Sao Paulo & bcc & Static & 106849 & 0.46 \\ \hline
		SK & Sao Paulo & bcc & WS & 106086 & -0.26 \\ \hline
		SK & Sao Paulo & fcc & Relaxed & 106223 & -0.13 \\ \hline
		SK & Sao Paulo & fcc & Static & 106748 & 0.37 \\ \hline
		SK & Sao Paulo & fcc & WS & 106386 & 0.02 \\ \hline
		SPVH & RMF & bcc & Static & 106692 & 0.31 \\ \hline
		SK & RMF & bcc & Relaxed & 105839 & -0.49 \\ \hline
		SK & RMF & bcc & Static & 106926 & 0.53 \\ \hline
		SK & RMF & bcc & WS & 105880 & -0.45 \\ \hline
		SK & RMF & fcc & Relaxed & 105913 & -0.42 \\ \hline
		SK & RMF & fcc & Static & 106749 & 0.37 \\ \hline
		SK & RMF & fcc & WS & 105925 & -0.41 \\ \hline
	\end{tabular}
  \end{minipage}
\end{table}

\chapter{CONTRIBUTORS FOR SOURCE CODE}
\label{appendix:contributions}

The original folding potential calculation software was written by myself for my Bachelor's of Science in Physics thesis from 2008.

Barbara Golf used this folding potential software as part of the calculation of the S-factor, and used that to calculate pycnonuclear reactions rates
for nuclei and strange quark matter in her thesis.
Whitney Ryan added the M3Y nucleon-nucleon interaction.
These capabilities were then retrofitted into \texttt{PycnoCalc} by me.

For this thesis I did the following:

\begin{description}[font=$\bullet$~\normalfont\scshape\color{red!50!black}]
\item Added the RMF nucleon-nucleon interaction
\item Added the temperature enhancements
\item Added the Schramm-Koonin formulation for reaction rates
\item Modified MESA to use my pycnonuclear reaction rates
\item Used Python Matplotlib and Pandas to produce charts
\end{description}

%
%
%
\addtocontents{toc}


\end{document}